%% file: IME-Paper2.tex
\documentclass[preprint,authoryear,12pt]{elsarticle}

\usepackage{graphicx}
\usepackage{stfloats}
\usepackage{caption}
\usepackage{subcaption}
\usepackage{tikz}

\usepackage{amsthm}
\usepackage{amsmath} 

\usepackage{verbatim}
\usepackage{cite}
\usepackage{epstopdf}
\usepackage{acronym}
\usepackage{ifthen}
\usepackage{multirow}

\journal{Insurance: Mathematics and Economics}

\begin{document}

\input{set_environment}

\def\bibsection{\section*{References}}

\begin{frontmatter}



\title{A Neural Network Approach to Efficient Valuation of Large Portfolios of Variable Annuities}


\author[CSUT]{Seyed Amir Hejazi}
\ead{amir@cs.toronto.edu}
\author[CSUT]{Kenneth R. Jackson}
\ead{krj@cs.toronto.edu}
\address[CSUT]{Department of Computer Science, University of Toronto, Toronto, ON, M5S 3G4, Canada}

\begin{abstract}
\input{abstract}
\end{abstract}

\begin{keyword}
Variable annuity \sep Spatial interpolation \sep Neural Network \sep Portfolio valuation
\end{keyword}

\end{frontmatter}



\input{introduction}

\input{literature_review}

\input{nn}

\input{numerical_experiments}
\input{conclusion}
\input{ack}



\appendix
\input{appendix}

\bibliographystyle{model2-names}
\bibliography{IME-Paper2}







\end{document}

%% file: set_environment.tex
\newtheorem*{quest}{Question}
\newtheorem{df}{Definition}
\newtheorem{eg}{Example}
\newtheorem{cor}{Corollary}
\newtheorem{thm}{Theorem}
\newtheorem{conj}{Conjecture}
\newtheorem{lem}{Lemma}

\newenvironment{ans}{\par\indent \emph{Ans:}}{\noindent \qedsymbol}    

\acrodef{VA}{Variable Annuity}
\acrodef{MC}{Monte Carlo}
\acrodef{IDW} {Inverse Distance Weighting}
\acrodef {RBF} {Radial Basis Function}
\acrodef {AV} {Account Value} 
\acrodef {GD} {Guaranteed Deat Benefit Value} 
\acrodef {CDF} {Cumulative Distribution Function}
\acrodef {LHS} {Latin Hypercube Sampling}
\acrodef {LSMC} {Least Squares Monte Carlo}
\acrodef {SCR} {Solvency Capital Requirement}
\acrodef {MSE} {Mean Squared Error}
\acrodef {NAG} {Nestrov's Accelerated Gradient}
\acrodef {RSS} {Replicated Stratified Sampling}

%% file: abstract.tex
Managing and hedging the risks associated with \ac{VA} products require 
intraday valuation of key risk metrics for these products. The complex structure of \ac{VA} 
products and computational complexity of their accurate evaluation have compelled 
insurance companies to adopt \ac{MC} simulations to value their large portfolios of \ac{VA} 
products. Because the \ac{MC} simulations are computationally demanding, especially for intraday 
valuations, insurance companies need more efficient valuation techniques.   
Recently, a framework based on traditional spatial interpolation techniques has been proposed that can 
significantly decrease the computational complexity of \ac{MC} simulation \citep{Gan15}.
However, traditional interpolation techniques require the definition of a distance function that can 
significantly impact their accuracy. Moreover, none of the traditional spatial interpolation techniques 
provide all of the key properties of accuracy, efficiency, and granularity \citep{Hejazi15}. In this paper, 
we present a neural network approach for the spatial interpolation framework that affords 
an efficient way to find an effective distance function. The proposed approach is accurate, efficient, and 
provides an accurate granular view of the input portfolio. Our numerical experiments illustrate the 
superiority of the performance of the proposed neural network approach compared to the traditional spatial 
interpolation schemes.   

%% file: introduction.tex
\section{Introduction}\label{sec:intro}
A Variable Annuity (VA), also known as a segregated fund in Canada, is a type of mutual fund that comes with insurance 
features and guarantees. \acp{VA} allow policyholders to invest in financial markets by making payment(s) 
into a predefined set of sub-accounts set 
up by insurance companies and enjoy tax-sheltered growth on their investment. The insurer, later, returns these 
investments through a lump-sum payment or a series of contractually agreed upon payments.
An attractive feature of \ac{VA} products are the embedded guarantees that protect the investment of policyholders 
from downside market fluctuations in a bear market and mortality risks \citep{Geneva13, Chi12}. 
For a detailed description of \ac{VA} products and different types of guarantees offered in these products, 
see our earlier paper \citep{Hejazi15} and the references therein. 

The innovative structure of embedded guarantees has made \ac{VA} products a huge success. Major insurance 
companies, especially in the past decade, have sold trillions of dollars worth of these products \citep{IRI11}, 
and have built up large portfolios of \ac{VA} contracts, each with hundreds of thousands of contracts.
The embedded guarantees of \ac{VA} contracts in these portfolios expose insurance companies to a substantial amount of 
risk, such as market risk and behavioral risk. After the market crash of 2008 that wiped out several big insurance 
companies, the surviving insurance companies started major risk management initiatives to dynamically hedge 
\citep{Hardy03} their exposures. 

\begin{figure}
\centering
\begin{tikzpicture}
 \fill[black] (1,3) circle (0.1cm);
 \foreach \y in {1,2,4}
 {
  \fill[black] (2, \y) circle (0.1cm);
  \draw[black,thick,->,>=latex] (1, 3) -- (1.9, \y);
  \foreach \yy in {-0.15,0,0.15,0.3}
   \draw[black,thick,->,>=latex] (2, \y) -- (3, \y + \yy);
 }
 \foreach \y in {2.75,3,3.25}
  \fill[black] (2, \y) circle (0.05cm);
\end{tikzpicture}
\caption{Pictorial graph of nested \ac{MC} simulations.}
\label{fig:nested-mc}
\end{figure}
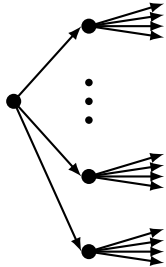

An integral part of the aforementioned hedging programs is intraday evaluation of \ac{VA} products to find 
the Greeks \citep{Hull06} for the portfolios of \ac{VA} products so that effective hedging positions can be set up.  
Most of the academic methodologies for valuation of \ac{VA} contracts are tailored to a specific type of 
\ac{VA} contract \citep{Milevsky06, Chen08-2, Chen08, Dai08, Ulm06, Huang11, Belanger09} 
and/or are computationally too expensive to scale to large portfolios of 
\ac{VA} contracts \citep{Azimzadeh15, Moenig11, Boyle08}. Hence, in practice, insurance companies have 
relied on nested \ac{MC} simulations to find the Greeks of \ac{VA} portfolios \citep{Reynolds08}. 
Nested \ac{MC} simulations, as shown in Figure \ref{fig:nested-mc}, consist of outer loop scenarios which span the space of 
key market variables and inner loop scenarios consisting of a collection of risk-neutral paths that are used to project 
the liabilities of \ac{VA} contracts \citep{Fox13}. Although \ac{MC} simulations are computationally less expensive than the 
academic methodologies, the amount of computation is still significant and does not scale well to 
large portfolios of \ac{VA} contracts. Because of this, insurance companies are actively looking for ways to 
reduce the number of required \ac{MC} simulations to find the Greeks for a large portfolio of \ac{VA} contracts.

As we discuss in Section \ref{sec:literature}, a framework based on spatial interpolation \citep{Burrough98} 
has been successful in ameliorating the computational load of \ac{MC} simulations by reducing the number of \ac{VA} contracts 
that go through nested \ac{MC} simulation. However, as we discussed in \citep{Hejazi15}, 
the proposed spatial interpolation framework requires an effective choice of distance function and a sample of \ac{VA} 
contracts from the space in which the input portfolio is defined to achieve an acceptable accuracy level. 
The appropriate choice of the distance function for the given input portfolio in the proposed framework requires 
research by a subject matter expert for 
the given input portfolio. In this paper, we propose to replace the conventional spatial interpolation techniques-- 
Kriging, \ac{IDW} and \ac{RBF} \citep{Burrough98}-- in the framework of \citep{Hejazi15} with a neural network. 
The proposed neural network can learn a good choice of distance function and use the given distance function to efficiently and 
accurately interpolate the Greeks for the input portfolio of \ac{VA} contracts. The proposed neural network only requires knowledge of 
a set of parameters that can fully describe the types of \ac{VA} contracts in the input portfolio and uses these parameters 
to find a good choice of distance function. 

The rest of this paper is organized as follows. Section \ref{sec:literature} provides a brief summary of existing 
methods for the valuation of portfolios of \ac{VA} products. The main focus of Section \ref{sec:literature} is 
on the spatial interpolation framework of \citep{Hejazi15} that has been successful in providing the Greeks for a large portfolio of 
\ac{VA} products in an efficient and accurate way. Section \ref{sec:nn} describes the neural network framework  
and provides background information on neural networks. We provide the intuition behind the proposed model and the 
novel training technique used to calibrate (a.k.a. to train) the network. Section \ref{sec:ne} provides insights into the performance of 
the neural network framework in estimation of Greeks for a large synthetic portfolio of \ac{VA} contracts. Section \ref{sec:conclusion} 
concludes the paper with a discussion of our future work and possible applications of the proposed framework.

%% file: literature_review.tex
\section{Portfolio Valuation Techniques}\label{sec:literature}
If one thinks of \acp{VA} as exotic market instruments \citep{Hull06}, the traditional replicating portfolio 
approach can be used to find the value of a portfolio of \ac{VA} products. The main idea behind this approach 
is to approximate the cash flow of liabilities for a portfolio of \ac{VA} contracts using well-formulated market 
instruments such as vanilla derivatives. The problem is often formulated as a convex optimization problem where the 
objective is to minimize the difference between the cash flow of the input portfolio and the replicating portfolio. 
Depending on the norm associated with the problem, linear programming \citep{Dembo99} or quadratic programming 
\citep{Daul09, Oechslin07} is used in the literature to find the replicating portfolio. The replicating portfolio, in our 
application of interest, doesn't provide us with an efficient alternative to \ac{MC} simulations, as one still needs to 
find the cash flow of the input portfolio for each year up to maturity. 

\ac{LSMC} regresses the liability of the input portfolio against some basis functions representing key economic 
factors \citep{Longstaff01, Carriere96}. \ac{LSMC} has been proposed in the literature to reduce the number of 
inner loop scenarios in nested \ac{MC} simulations \citep{Cathcart09}. Depending on the type of embedded 
guarantees, size of investment and characteristics of the policyholder, \ac{VA} contracts have a significant 
number of numeric attributes, each covering a broad range. Therefore, an accurate regression using \ac{LSMC} 
requires incorporation of many sample points, and hence is computationally demanding. 

Recently, \ac{RSS} \citep{Vadiveloo11} and Kriging based techniques \citep{Gan13-2, Gan15} 
have been proposed to reduce the number of \ac{VA} contracts that must be included in the \ac{MC} simulations. 
Both of these methods, use the Greeks for samples of the input portfolio to estimate the Greeks of the full input portfolio. 
\ac{RSS} requires several iterations of sample generation and evaluation to converge to a final result. 
This makes it more computationally demanding than the Kriging based techniques of \citep{Gan13-2, Gan15} that require 
\ac{MC} simulations results for only one sample. We discuss in our earlier paper \citep{Hejazi15} how the Kriging 
based techniques of \citep{Gan13-2, Gan15} can be categorized under a general spatial interpolation framework. 
The spatial interpolation framework generates a sample of \ac{VA} contracts from the space in which the \ac{VA} 
contracts of the input portfolio are defined. The Greeks for the sample are evaluated using nested \ac{MC} simulations. 
The results of \ac{MC} simulations are then used by a spatial interpolation technique to generate an estimate 
for the Greeks of the input portfolio. 

In \citep{Hejazi15}, we provide numerical and theoretical results comparing the efficiency and accuracy of different conventional 
spatial interpolation techniques, i.e., Kriging, \ac{IDW} and \ac{RBF}. Our results demonstrate that, while the Kriging method provides 
better accuracy than either the \ac{IDW} method or the \ac{RBF} method, it is less efficient and has a lower resolution. 
By lower resolution, we mean that the Kriging method can provide the Greeks for only the input portfolio in an efficient manner, 
while both the \ac{IDW} method and the \ac{RBF} method approximate the Greeks efficiently for each \ac{VA} contract in the input portfolio.

%% file: nn.tex
\section{Neural Network Framework}\label{sec:nn}
As we discuss in our earlier paper \citep{Hejazi15}, spatial interpolation techniques 
can provide efficient and accurate estimation of the Greeks for a large portfolio of \ac{VA} 
products. Although \ac{IDW} and \ac{RBF}
methods provide better efficiency and resolution than Kriging methods, they are less accurate than 
Kriging methods. Our experiments in \citep{Hejazi15} demonstrate the significance of the choice of 
distance function on the accuracy of \ac{IDW} and \ac{RBF} methods. A manual approach to find the best 
distance function that minimizes the estimation error of the \ac{IDW} and the \ac{RBF} methods for a 
given set of input data is not straightforward and requires investing a significant amount of time. 
The difficulty in finding a good distance function diminishes the effectiveness of the \ac{IDW} and the \ac{RBF} methods. 

In order to automate our search for an effective distance function while maintaining the 
efficiency of the \ac{IDW} and the \ac{RBF} methods, we propose a machine learning approach. 
In our proposed approach, we use an extended version of the Nadaraya-Watson kernel regression 
model \citep{Nadaraya64, Watson64} to estimate the Greeks. Assuming $y(z_1), \cdots, y(z_n)$ are 
the observed values at known locations $z_1, \cdots, z_n$, the Nadaraya-Watson estimator 
approximates the value $y(z)$ at the location $z$ by

\begin{equation*}\label{eq:nw-model}
\hat{y}(z) = \sum_{i = 1}^n \frac{K_h(z - z_i) \times y(z_i)}{\sum_{j = 1}^n K_h(z - z_j)}
\end{equation*} 
where $K_h$ is a kernel with a bandwidth of $h$. The Nadaraya-Watson estimator was first proposed for 
kernel regression applications and hence the choice of kernel function $K_h$ was a necessity. For our 
application of interest, we choose to use the following extended version of the Nadaraya-Watson estimator: 

\begin{equation}\label{eq:est-model}
\hat{y}(z) = \sum_{i = 1}^n \frac{G_{h_i}(z - z_i) \times y(z_i)}{\sum_{j = 1}^n G_{h_j}(z - z_j)}
\end{equation}  
where $G$ is a nonlinear differentiable function and the subscript $h_i$, similar to the bandwidth $h$ 
of kernels, denotes the range of influence of each $y(z_i)$ on the estimated value. Unlike the Nadaraya-Watson 
model, the $h_i$s are not universal free parameters and are location dependent. Moreover, $h_i$ is a vector 
that determines the range of influence of each pointwise estimator in each direction of feature space of 
the input data.  As we discuss below, our decision to calibrate the $h_i$ parameters using a neural network 
necessitated the properties of $G$.     

In our application of interest, the $z_i, 1 \leq i \leq n$, in \eqref{eq:est-model} define a set of \ac{VA} contracts, 
called representative contracts, and $y(z_i), 1\leq i \leq n$, are their corresponding Greek values. Hence, 
Equation \eqref{eq:est-model} is similar to the equation for the \ac{IDW} estimator. The $G_{h_i}(z - z_i)$ in 
\eqref{eq:est-model} is comparable to the weight (inverse of the distance) for representative contract $z_i$ 
in the equation of \ac{IDW}. Therefore, once we know the particular choices of the $h_i$s for the $G$ function 
for each of our $n$ representative \ac{VA} contracts, we can compute the Greeks for a large portfolio of $N$ 
\ac{VA} contracts in time proportional to $N \times n$, which preserves the efficiency of our framework. 
In order to find a good choice of the $G_{h_i}$ functions, we propose the use of a particular type of neural network called 
a feed-forward neural network. As we describe in more detail below, our choice of neural network allows 
us to find an effective choice of the $G_{h_i}$ functions by finding the optimum choices of the $h_i$ values that minimize 
our estimation error, and hence eliminate the need for a manual search of a good choice of distance function. 

Feed-forward networks are well-known for their general approximation properties which has 
given them the name of universal approximators. For example a one-layer feed-forward network with 
linear outputs is capable of approximating any continuous function on a compact domain \citep{Hornik91}. For a thorough 
study of feed-forward networks, the interested reader is referred to \citep{Bishop06} and the references therein. 
For the sake of brevity, in the rest of this paper, we use the word neural network to refer to this particular 
class of feed-forward neural network unless explicitly said otherwise.

\subsection{The Neural Network}
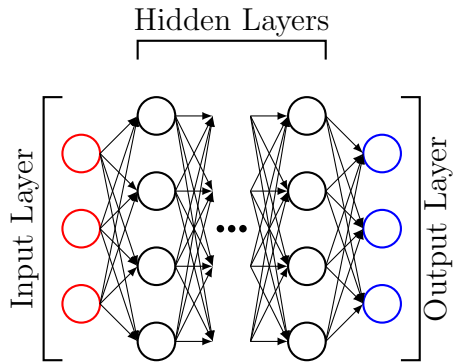
\begin{figure}[bt]
\centering
\begin{tikzpicture}
 \foreach \y in {1,2,3}
  \draw[red, thick] (1,\y + 0.5) circle (0.25cm);
 \foreach \y in {1,...,4}
  \draw[black, thick] (2,\y) circle (0.25cm);
 \foreach \x in {2.85,3,3.15}
  \fill[black] (\x, 2.5) circle(0.05cm);
 \foreach \y in {1,...,4}
  \draw[black, thick] (4,\y) circle (0.25cm);
 \foreach \y in {1,2,3}
  \draw[blue, thick] (5,\y + 0.5) circle (0.25cm);
 \foreach \y in {1,2,3}
  \foreach \yy in {1,...,4}
   \draw[black,->,>=latex] (1.25,\y + 0.5) -- (1.75, \yy);

 \foreach \y in {1,...,4}
  \foreach \yy in {1,...,4}
  {
   \draw[black,->,>=latex] (2.25,\y) -- (2.75, \yy);
   \draw[black,->,>=latex] (3.25,\y) -- (3.75,\yy);
  }

 \foreach \y in {1,...,4}
  \foreach \yy in {1,2,3}
   \draw[black,->,>=latex] (4.25,\y) -- (4.75, \yy + 0.5);
  
 \draw[black, thick] (0.75,0.75) -- (0.5,0.75) -- (0.5,4.25) -- (0.75,4.25);
 \node[rotate=90] at (0.25, 2.5){Input Layer};
 \draw[black, thick] (1.75,4.75) -- (1.75,5) -- (4.25,5) -- (4.25,4.75);
 \node at (3,5.25){Hidden Layers};
 \draw[black, thick] (5.25,4.25) -- (5.5, 4.25) -- (5.5,0.75) -- (5.25,0.75);
 \node[rotate=90] at (5.75, 2.5){Output Layer};
\end{tikzpicture}
\caption{Diagram of a feed-forward neural network. Each circle represents a neuron.}
\label{fig:nn}
\end{figure}

A feed-forward neural network is a collection 
of interconnected processing units, called neurons, which are organized in different layers (Figure \ref{fig:nn}). 
The first and the last layers are respectively called the input layer and the output layer. 
Intermediate layers are called the hidden layers. Neurons of each layer take as input the outputs 
of the neurons in the previous layer. The neurons in the first layer serve only as inputs to 
the network. In other words, the neurons of the input layer produce what is known as the feature vector.

Assuming $x_1, \cdots, x_n$ are the inputs of neuron $j$ at hidden level $l$. First a linear 
combination of input variables is constructed at each neuron: 

\begin{equation*}\label{eq:activation}
a^{(l)}_j = \sum_{i = 1}^{n} w^{(l)}_{ij}x_i + b^{(l)}_j
\end{equation*}
where parameters $w_{ij}$ are referred to as weights and parameter $b_j$ is  called 
the bias. The quantity $a_j^{(l)}$ is known as the activation of neuron $j$ at level $l$. 
The activation $a_j^{(l)}$ is then transformed using a differentiable, nonlinear function to 
give the output of neuron $j$ at level $l$.

\begin{figure}[bt]
\centering
\begin{tikzpicture}
 \foreach \y in {2,4,8}
  \draw[red, thick] (2.25,\y - 0.75) rectangle (2.75, \y + 0.75);
 \foreach \y in {5.5,6,6.5}
  \fill[black] (2.5, \y) circle(0.05cm);
  
 \foreach \y in {2,4,8}
  \draw[black, thick] (4,\y) circle (0.25cm);
 \foreach \y in {5.5,6,6.5}
  \fill[black] (4, \y) circle(0.05cm);

 \foreach \y in {2,4,8}
  \draw[blue, thick] (6,\y) circle (0.25cm);
 \foreach \y in {5.5,6,6.5}
  \fill[black] (6, \y) circle(0.05cm);

 \foreach \y in {2,4,8}
  \foreach \yy in {2,4,8}
  {
   \draw[black,->,>=latex] (2.75,\y - 0.4) -- (3.75, \y - 0.1);
   \draw[black,->,>=latex] (2.75,\y) -- (3.75, \y);
   \draw[black,->,>=latex] (2.75,\y + 0.4) -- (3.75, \y + 0.1);
  }

 \foreach \y in {2,4,8}
  \foreach \yy in {2,4,8}
   \draw[black,->,>=latex] (4.25,\y) -- (5.75, \yy);

 \draw[black, thick] (0.5,1.75) -- (0.25,1.75) -- (0.25,8.25) -- (0.5,8.25);
 \node[rotate=90] at (0, 5){Input Layer};
 \node[] at (1.25, 2){\begin{tabular}{c}$z_1$\\ Features\end{tabular}};
 \node[] at (1.25, 4){\begin{tabular}{c}$z_2$\\ Features\end{tabular}};
 \node[] at (1.25, 8){\begin{tabular}{c}$z_n$\\ Features\end{tabular}};

 \draw[black, thick] (3,9) -- (3,9.25) -- (5,9.25) -- (5,9);
 \node at (4,9.5){Hidden Layer};
 \draw[black, thick] (6.25,8.25) -- (6.5, 8.25) -- (6.5,1.75) -- (6.25,1.75);
 \node[rotate=90] at (6.75, 5){Output Layer};
\end{tikzpicture}
\caption{Diagram of the proposed neural network. Each circle represents a neuron. Each rectangle represent the 
set of neurons that contains input features corresponding to a representative contract.}
\label{fig:ddnet}
\end{figure}
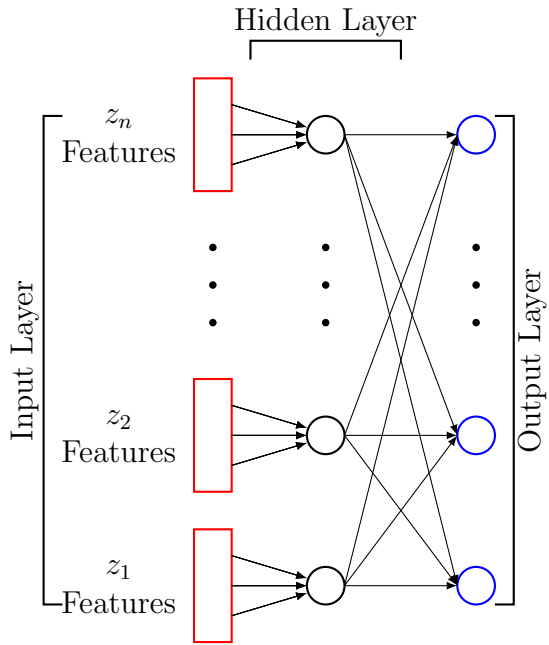

In our framework, we propose to use a neural network with only one hidden layer (Figure \ref{fig:ddnet}). 
Each neuron in the input layer represents a value in the set $\{F^c, F^-, F^+\}$. 
Each $f$ in $F^c$ assumes the following form 

\begin{equation*}
 f = \Big\{\begin{matrix} 0 & \text{if }x_c = x_{c_i}\\ 1 & \text{if }x_c \neq x_{c_i}\end{matrix}
\end{equation*}
where $x_c$ represents the category of categorical attribute $c$ for input \ac{VA} policy $z$, 
and $x_{c_i}$ represents the category of categorical attribute $c$ for representative \ac{VA} policy 
$z_i$ in the sample.
Each value $f$ in $F^-$ has the form $f = [t(\mathbf{x}_{n_i}) - t(\mathbf{x}_n)]^+/R_{t_h}$, and each 
value $f$ in $F^+$ has the form $f = [t(\mathbf{x}_n) - t(\mathbf{x}_{n_i})]^+/R_{t_h}$. In both of the aforementioned 
formulas, $\mathbf{x}_n$ is the vector containing the numeric attributes of input \ac{VA} policy $z$, 
$\mathbf{x}_{n_i}$ is the vector containing the numeric attributes of representative \ac{VA} policy $z_i$ 
in the sample, $t(\cdot)$ is a transformation (linear/nonlinear), determined by the expert user, 
that assumes a value in an interval of length $R_{t}$ and $[\cdot]^+ = \max(\cdot, 0)$. 
In essence, our choice of input values allows different bandwidths 
($h_i$ values in \eqref{eq:est-model}) to be used for different attributes of \ac{VA} policies and in 
different directions around a representative \ac{VA} contract in the sample.
Since we are interested in calibrating the $G$ functions of equation \eqref{eq:est-model}, the 
number of neurons in the output and hidden layer equals the number of representative 
contracts in the sample. The inputs of neuron $i$ in the hidden layer are those values of $f$ 
in the input layer that are related 
to the representative \ac{VA} policy $i$. In other words, input values of neuron $i$ in the hidden layer determine 
the per attribute difference of input \ac{VA} contract $z$ with the representative \ac{VA} contract 
$z_i$ using the features $f \in \{F^c, F^-, F^+\}$.
Each neuron of the hidden layer transforms its activation using an exponential function 
to form its output. The output of neuron $i$ in the output layer is the normalized version of the output for  
neuron $i$ in the hidden layer. Hence the outputs of the network, i.e., $o_i, i \in \{1,\cdots,n\}$, in machine 
learning terminology, correspond to a softmax of activations in the hidden layer. These outputs can be used 
to estimate the value of the Greek for input \ac{VA} $z$ as $\hat{y}(z) = \sum_{i = 1}^n o_i \times y(z_i)$, 
in which $y(z_i)$ is the value of the Greek for sample \ac{VA} policy $z_i$. To summarize, our proposed neural 
network allows us to rewrite Equation \eqref{eq:est-model} as 

\begin{equation}\label{eq:nn-model}
\hat{y}(z) = \sum_{i = 1}^n \frac{\exp(\mathbf{\mathbf{w}_i}^T \mathbf{f}(z, z_i) + b_i) \times y(z_i)}{\sum_{j = 1}^n \exp(\mathbf{\mathbf{w}_j }^T \mathbf{f}(z, z_j) + b_j)}
\end{equation}
where vector $\mathbf{f}(z, z_i)$ represents the features in the input layer that are related to the representative 
\ac{VA} policy $z_i$, and vector $\mathbf{w_i}$ contains the weights associated with each feature in 
$\mathbf{f}$ at neuron $i$ of the hidden layer. Each $\mathbf{w_i}$ in \eqref{eq:nn-model} can be 
considered as the pointwise inverse of the bandwidth value $h_i$ in \eqref{eq:est-model}.

\subsection{Network Training Methodology}\label{sec:net-train}
Equation \eqref{eq:nn-model} is a parametric formulation of our proposed estimator. 
We have to calibrate the weights and bias parameters to find an estimator with minimum estimation error. 
The calibration process, in neural network literature, is known as network training. 
In order to minimize the training time by reducing the number of \ac{VA} policies for which the 
framework has to do \ac{MC} estimations, we select a small set of \ac{VA} policies which we call the 
training portfolio as the training data for the network.  
The objective of the calibration process is then to find a set of weights and bias parameters that minimizes the 
\ac{MSE} in the estimation of the Greeks of the training portfolio.

We choose the training portfolio to be different 
than the set of representative \ac{VA} policies (i.e., observed points in the model \eqref{eq:est-model}) to 
avoid data overfitting. Even with this choice of the training data, one cannot avoid the issue of overfitting. We 
discuss in Section \ref{sec:stop-cond} our solution to this problem. 

Following the common practice in neural network literature, we use the following simple 
gradient descent scheme \citep{Boyd04} to iteratively update the weight and bias parameters. 

\begin{equation}\label{eq:gradient-descent}
[\mathbf{w}^{(t + 1)}, \mathbf{b}^{(t + 1)}] = [\mathbf{w}^{(t)}, \mathbf{b}^{(t)}] - \eta \nabla E(\mathbf{w}^{(t)}, \mathbf{b}^{(t)})
\end{equation}
The parameter $\eta > 0$ in \eqref{eq:gradient-descent} is known as the learning rate and  
$\mathbf{w}^{(t)}$ and $\mathbf{b}^{(t)}$ denote the vectors of the network's weights and biases, 
respectively, at iteration $t$. $E(\cdot)$ represents the error function that 
we are trying to minimize and $\nabla E$ is the gradient of $E$. 

For a fair comparison with the traditional spatial interpolation techniques discussed in \citep{Hejazi15}, 
training the network to minimize the following \ac{MSE} in estimation of the Greeks for the training portfolio 
seems to be a reasonable approach. 

\begin{equation}\label{eq:mse-error}
E(\mathbf{w}, \mathbf{b}) = \frac{1}{2|B|} \sum_{k = 1}^{|B|} || \hat{y}(\bar{z}_k, \mathbf{w}, \mathbf{b}) - y(\bar{z}_k)||^2
\end{equation}
where $\bar{z}_k, 1 \leq k \leq |B|$, are the \ac{VA} policies in the training portfolio.

Depending on the application of interest, the $y(\bar{z}_i)$ values can be too small (too big) resulting in too small 
(too big) gradient values for \eqref{eq:mse-error}. Too small gradient values increase the training time to 
reach a local minimum, while too big gradient values cause big jumps in updates of \eqref{eq:gradient-descent} 
and hence numerical instability. Normalizing the values of $y(\bar{z}_i)$ in \eqref{eq:mse-error} and 
the choice of learning rate can help to ameliorate this problem. 

Currently, in the field of machine learning, there does not exist a universal objective approach for choosing the values 
of the free parameters, such as the learning rate, that we describe in this section. 
Choosing appropriate values of almost all of these free parameters requires subjective judgement and is dependent on the
data that is used and the application of interest. Although there exist many theoretical results \citep{Boyd04} that show
that a particular choice of these parameters, e.g., learning rate, can guarantee bounds on the convergence rate of the 
training method and these results are helpful in setting the theoretical framework for the training method, 
experimentation and subjective judgement must still be used to determine the actual values of many of these parameters.  
For example, for many problems, it is possible to prove that, if the the learning rate is sufficiently small, 
then the gradient descent method \eqref{eq:gradient-descent} will converge to a local minimum.  Unfortunately, 
this result, or similar well-known results in the optimization or machine learning literature, do not say how to 
choose a ``good'' value for the learning rate that will ensure that the gradient descent method \eqref{eq:gradient-descent} 
yields near optimal values of the neural network parameters with an amount of computational work that is close to minimal. 
However, there is a well-established heuristic approach on how to choose the free parameters to meet this goal and/or 
to avoid problems such as overfitting (see, for example, \citep{Murphy12, Bishop06} and references there in). 
Based on these guidelines, in Appendix A, we explain some simple heuristics that allow one easily to choose 
the free parameters described in this section.

Our formulation of error function \eqref{eq:mse-error} uses the whole training set to 
compute the error function and subsequently the gradient of the error function in each iteration. 
Training techniques that use the whole training set in each iteration are known as batch methods \citep{Bishop06}. 
Because of the redundancy in the data as well as the computational complexity of evaluating gradients, batch 
gradient descent is a slow algorithm for training the network. Our experiments, further, corroborated the slowness 
of batch gradient descent in training our proposed network. To speed up the training, we used a particular version of what is 
known as the mini-batch training method \citep{Murphy12}. In our training method, in each iteration, we select 
a small number ($\leq 20$) of training \ac{VA} policies at random and train the network using the gradient of the 
error function for this batch. Hence, the error function in our mini-batch training method has the form

\begin{equation}\label{eq:mini-batch-error}
E(\mathbf{w^{(t)}}, \mathbf{b^{(t)}}) = \frac{1}{2|B^{(t)}|} \sum_{k \in B^{(t)}} || \hat{y}(\bar{z}_k, \mathbf{w^{(t)}}, \mathbf{b^{(t)}}) - y(\bar{z}_k)||^2
\end{equation}
where $B^{(t)}$ is the set of indices for selected \ac{VA} policies at iteration $t$.

Gradient descent methods, at each iteration, produce a higher rate of reduction in the directions of high-curvature than 
in the directions of lower-curvature. Big rate reductions in directions of high-curvature cause zig-zag 
movements around a path that converges to the local minimum and hence decrease the convergence rate \citep{Murphy12}. However, 
a slower rate of reduction in directions of low-curvature allows for a persistent movement along the path of convergence to 
the local minimum. We can exploit this property by changing the weight update policy of gradient descent to use a 
velocity vector that increases in value in the direction of persistent reduction in the objective error function 
across iterations. This techniques is known as the momentum method \citep{Polyak64}. In our training method, we use 
\ac{NAG} method \citep{Nesterov83} which can be considered as a variant of the classical momentum method \citep{Polyak64}. 
In particular, we use a version of the \ac{NAG} method described in \citep{Sutskever13} in which 
the \ac{NAG} updates can be written as 

\begin{align*}\label{eq:momentum}
&v_{t+1} = \mu_t v_t - \epsilon \nabla E([\mathbf{w}^{(t)}, \mathbf{b}^{(t)}] + \mu_t v_t) \nonumber \\
&[\mathbf{w}^{(t + 1)}, \mathbf{b}^{(t + 1)}] = [\mathbf{w}^{(t)}, \mathbf{b}^{(t)}] + v_{t + 1}
\end{align*}
where $v_t$ is the velocity vector, $\mu_t \in [0, 1]$ is known as the momentum coefficient and $\epsilon$ is the 
learning rate. In this scheme, the momentum coefficient is an adaptive parameter defined by 

\begin{equation}\label{eq:momentum-coeff}
\mu_t = \min (1 - 2^{-1 - \log_2(\lfloor \frac{t}{50} \rfloor + 1)}, \mu_{\max})
\end{equation}  
where $\mu_{\max} \in [0, 1]$ is a user defined constant. For general smooth 
convex functions and a deterministic gradient, \ac{NAG} achieves a global convergence rate of $O(\frac{1}{T^2})$
\footnote{f(x) = O(g(x)) means that there exist positive numbers $c$ and $M$ such that $\forall x > M: f(x) < cg(x)$} 
versus the $O(\frac{1}{T})$ convergence rate for gradient descent in which $T$ denotes the number of iterations \citep{Sutskever13}. 
In this context, the rate of convergence is defined as the rate at which the error, $|f(x_T) - f(x^*)|$, goes to zero, 
where $f$ is a smooth convex function, $x^*$ is the optimum value (value of interest) and $x_T$ is the estimation 
of $x^*$ after iteration $T$. 

\subsection{Stopping Condition}\label{sec:stop-cond}
\begin{figure}[bt]
\centering
\includegraphics[width=\textwidth, height=0.5\textheight]{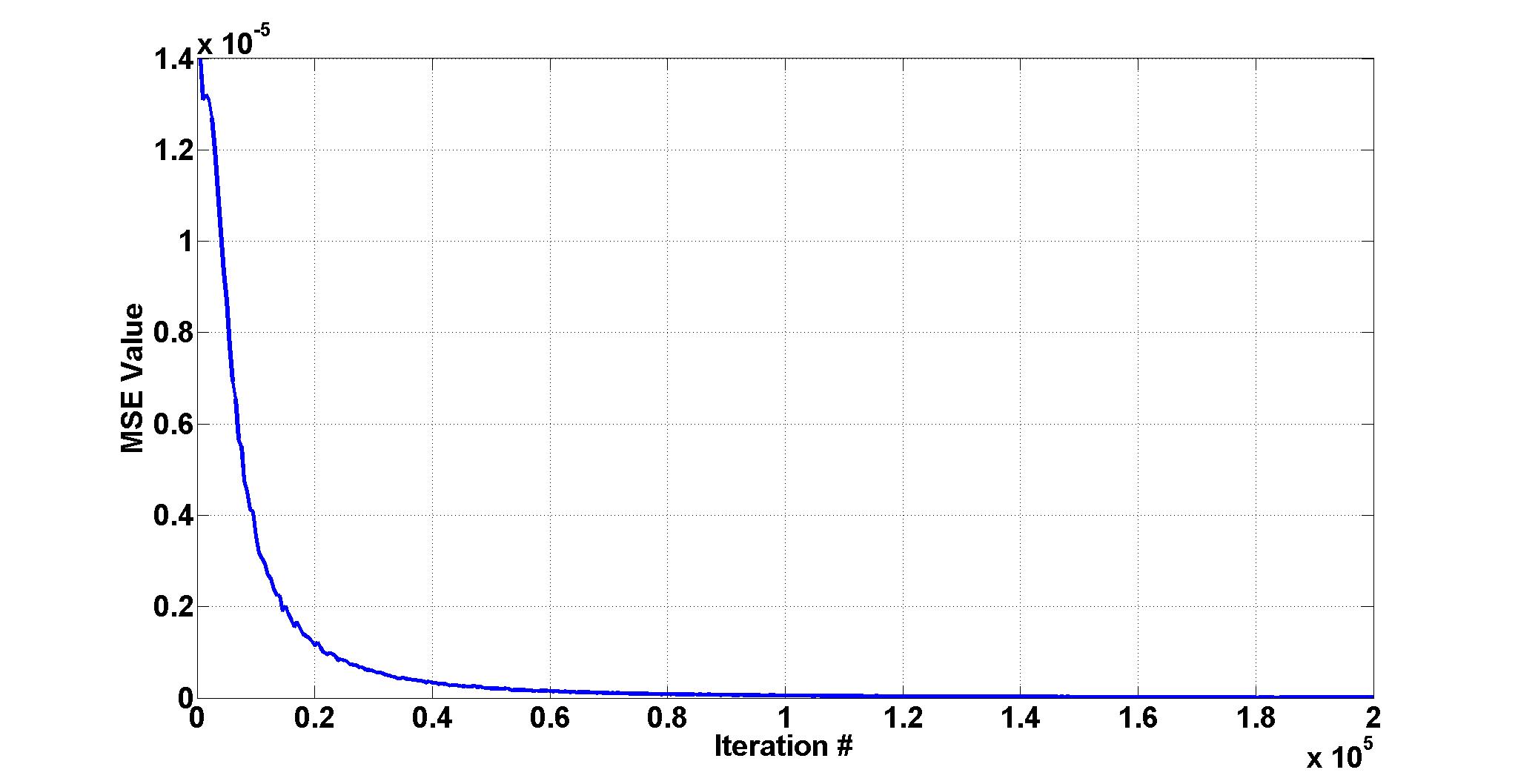}
\caption{MSE of \ac{VA} policies in the batch as a function of the iteration number.}
\label{fig:mse-iter}
\end{figure}

Figure \ref{fig:mse-iter} represents a graph of the \ac{MSE} for a set of training \ac{VA} policies as a function 
of the training iteration number for one run of our training algorithm. The graph, except at a few points, is a 
decreasing function of the iteration number which means that, as the iteration proceeds, the network is learning 
and steadily improving the bandwidth parameters for the model \eqref{eq:nn-model}. 
After the first few thousand iterations, the graph of Figure \ref{fig:mse-iter} kneels and the 
rate of decrease in \ac{MSE} drops dramatically. Such significant drops in the rate of \ac{MSE} reduction 
is a sign that the network parameters are close to their respective optimum values. If we 
train the network for a longer time, we expect the \ac{MSE} to continue to decrease slowly. However, the amount of 
improvement in the accuracy of the network might not be worth the time that we spend in further training the network. 
Hence, it might be best to stop the training. 

\begin{figure}[bt]
\centering
\includegraphics[width=\textwidth, height=0.5\textheight]{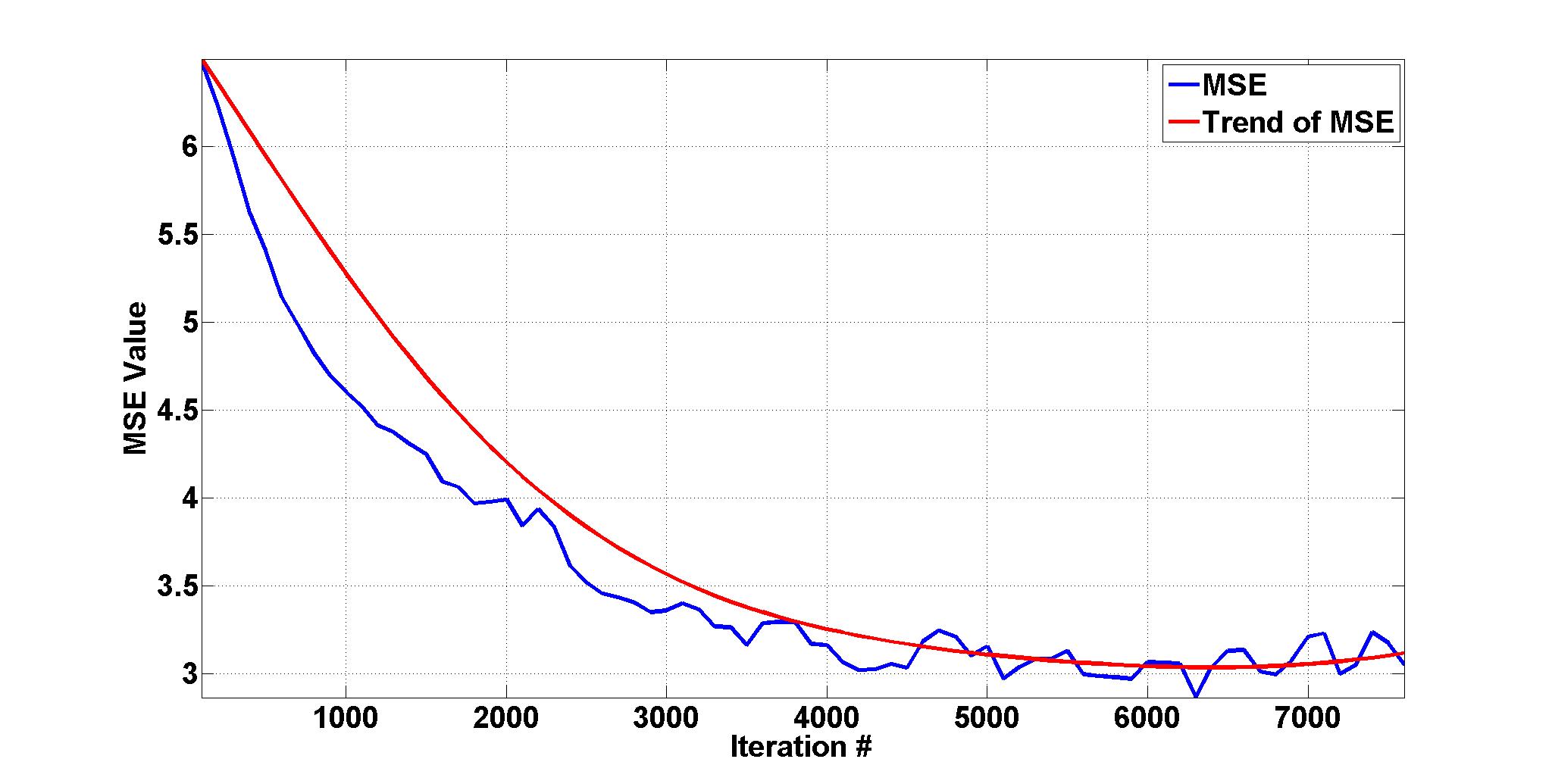}
\caption{The \ac{MSE} of the validation set and the trend in the \ac{MSE} as a function of the iteration number for a run of the training algorithm. 
The trend is found using a moving average with a window size of 10 and then fitting a polynomial of degree 6 to the smoothed data.}
\label{fig:validation-err}
\end{figure}

If we select \ac{VA} policies for the training portfolio very close to the representative \ac{VA} policies, 
training the network for a long time can cause data overfitting. Because a perfect solution for 
\eqref{eq:nn-model}, in this case, is achieved when the bandwidth values tend to zero or equivalently the 
weight parameters become very large. However, such a network approximates the Greeks of \ac{VA} policies that 
are not close to the representative \ac{VA} policies by zero. To avoid 
over-training the network in such scenarios, we follow the common practice in the machine learning literature and 
track the \ac{MSE} for a set of \ac{VA} policies which we call the validation portfolio \citep{Murphy12}. 
The validation portfolio is a small set of \ac{VA} policies that are selected uniformly at random from the \ac{VA} 
policies in the input portfolio. The \ac{MSE} of the validation set should decrease at first as the network 
learns optimal parameters for the model \eqref{eq:nn-model}. After reaching a minimum value, the \ac{MSE} of the validation 
portfolio often increases as the network starts to overfit the model \eqref{eq:nn-model} 
(Figure \ref{fig:validation-err}). In our training method, we propose to evaluate 
the \ac{MSE} of the validation portfolio every $I^{th}$ iteration of training, to avoid significantly slowing down 
the training process. We also propose to  use a window of length $W$ of the recorded \ac{MSE} values for the validation set to 
determine if the \ac{MSE} of the validation set has increased in the past $W - 1$ recorded values after attaining a 
minimum. If we find such a trend, we stop the training to avoid overfitting. 
$I$ and $W$ are user defined (free) parameters and are application dependent. 

As shown in the graph of Figure \ref{fig:validation-err}, the actual graph of the \ac{MSE} for the validation portfolio as a function 
of iteration number might be volatile. However, a general u-shaped trend still exists in the data, which illustrates an increase 
in the value of the \ac{MSE} after the \ac{MSE} has reached a minimum. In order to find the trend graph, we use a simple 
moving average with a window size of $\bar{W}$ to smooth the data. We then fit, in the \ac{MSE} sense, a polynomial of degree $d$ to 
the smoothed data. We examine the resulting trend graph with windows 
of length $W$ to determine the phenomenon of the \ac{MSE} increase after attaining a local minimum. The parameters $\bar{W}$ and $d$ 
are free parameters and are dependent on the application of interest. 

So far, we have discussed two events, which we call stopping events, that can be used as indicators to stop the training. 
In both events, the network parameters are close to optimal network parameter values. At this point, each additional 
iteration of the training algorithm moves these parameters in a neighborhood of the optimal values and might make the 
network parameters closer to the optimal values or farther from the optimal values. 
Intuitively, the best time to stop the training is when the network parameters are very close to the optimal values and 
further training does not significantly improve them. In our training 
algorithm, we propose to use the relative error in an estimation of the mean of the Greeks of the validation portfolio as 
our stopping criteria. Let $\bar{G_{NN}}$ and $\bar{G_{MC}}$ denote the mean of the estimated Greeks for the validation portfolio 
computed by our proposed neural network approach and by \ac{MC} simulations respectively. The relative error in estimation of the mean of the 
Greeks for the validation portfolio is then 

\begin{equation}\label{eq:G-rel-err}
\text{Err} = \frac{|\bar{G_{NN}} - \bar{G_{MC}}|}{|\bar{G_{MC}}|}
\end{equation}
If the relative error \eqref{eq:G-rel-err} is smaller than a user defined threshold $\delta$, we stop the training. 
The idea behind our choice of stopping criteria is that a good validation portfolio should be a good representative of 
the input portfolio. Hence, a network that has, on average, an acceptable accuracy in an estimation of the Greeks 
for the validation portfolio should, intuitively, have an acceptable accuracy in an estimation of the Greeks for the 
input portfolio as well. In some cases, finding stopping events and satisfying the stopping criteria may require the training 
algorithm to go through too many iterations, which can significantly increase the training time of the network and consequently 
decrease the efficiency of the method. We propose to stop the training algorithm once the network has gone through a user defined 
maximum number of iterations to limit the training time in such scenarios. 

\subsection{Sampling}
As we discuss in our earlier paper \citep{Hejazi15}, the choice of an appropriate sampling method is a key factor in obtaining 
an effective method within the proposed spatial interpolation framework. Although we do not address the issue of selecting an 
effective sampling method in this paper, in this section, we describe ways in which the choice of our representative \ac{VA} 
contracts can affect the performance of our proposed method.

Consider a realization of our proposed network with three representative contracts $x_1, x_2^{(1)}$ and $x_3$ with similar guarantee types. 
The \ac{VA} contracts $x_1$ and $x_2^{(1)}$ are similar in every attribute except for the numeric attribute $a_n$ and they differ with \ac{VA} 
contract $x_3$ in every attribute. Now, consider another realization of our proposed network in which we replace $x_2^{(1)}$ in the aforementioned 
realization with $x_2^{(2)}$. We choose $x_2^{(2)}$ such that it has similar categorical attributes as $x_2^{(1)}$; however, its numeric attributes 
assume the average of the corresponding numeric values for $x_1$ and $x_3$. Assume we train both networks for a similar number of iterations $I$.
The gradient values of the error function depend only on the network architecture and the choice of input values. Since the input values 
for the corresponding hidden layer neurons for $x_1$ and $x_2^{(1)}$ in the former network are almost equal we expect the corresponding weight vectors 
$\mathbf{w_1^{(1)}}$ and $\mathbf{w_2^{(1)}}$ for these neurons to be approximately equal as well. However, because of the dissimilarity of the 
$x_1$ and $x_2^{(2)}$ contracts in the second network, we expect the input values and hence the corresponding weights $\mathbf{w_1^{(2)}}$ and $\mathbf{w_2^{(2)}}$ 
of the hidden layer neurons corresponding to these contracts to be quite different. Consequently, the latter network can provide a better differentiation 
between the $x_1$ and $x_2^{(2)}$ contracts while the former network requires more training time to provide the same level of accuracy in differentiating 
$x_1$ and $x_2^{(1)}$. Moreover, in approximating the Greeks for \ac{VA} contracts other than $x_1, x_2^{(1)}, x_2^{(2)}$ and $x_3$, the former 
network, because of the similarity in weights $\mathbf{w_1^{(1)}}$ and $\mathbf{w_2^{(1)}}$, puts more emphasis on the corresponding Greeks of the contracts $x_1$ 
and $x_2^{(1)}$. Moreover, the latter network, because of the choice of $x_2^{(2)}$, can provide better accuracy for \ac{VA} contracts that 
are quite different than both $x_1$ and $x_3$. Therefore, as demonstrated by this example, a bad sample can hurt the efficiency of the proposed 
method by requiring more training time. Moreover, a bad sample can hurt the accuracy of the proposed network in estimation of the Greeks 
of \ac{VA} contracts that assume attribute values that are different than the representative contracts, in particular those \ac{VA} contracts that 
are quite distant from any representative contract.

%% file: numerical_experiments.tex
\section{Numerical Experiments}\label{sec:ne}
In this section, we provide numerical results illustrating the performance 
of the proposed neural network framework in comparison with the traditional 
spatial interpolation schemes (i.e., Kriging, \ac{IDW}, and \ac{RBF}) discussed in \citep{Hejazi15}. 
The input portfolio in all experiments is a synthetic portfolio of $100,000$ 
\ac{VA} contracts with attribute values that are chosen uniformly at random from 
the space described in Table \ref{tb:portfolio}. Similar to \citep{Hejazi15}, we 
allow guarantee values to be different than the account values. The guaranteed 
death benefit of contracts with a GMWB rider is set to be equal to their guaranteed 
withdrawal benefit. The account values of the contracts follow a simple log-normal 
distribution model \citep{Hull06} with a risk free rate of return of $\mu = 3\%$, 
and volatility of $\sigma = 20\%$.  

\begin{table}[!bt]
 \centering
 \begin{tabular}{|l|l|}
  \hline
  {\bf Attribute} & {\bf Value}\\ 
  \hline
  Guarantee Type & \{GMDB, GMDB + GMWB\} \\
  \hline
  Gender & \{Male, Female\}\\
  \hline
  Age & $\{20, 21, \ldots, 60\}$\\
  \hline
  Account Value & $[1e4, 5e5]$ \\ 
  \hline 
  Guarantee Value & $[0.5e4, 6e5]$ \\
  \hline
  Withdrawal Rate & $\{0.04, 0.05, 0.06, 0.07, 0.08\}$\\
  \hline
  Maturity & $\{10, 11, \ldots, 25\}$\\
  \hline
 \end{tabular}
 \caption{GMDB and GMWB attributes and their respective ranges of values.}
 \label{tb:portfolio}
\end{table}

In our experiments, we use the framework described in \citep{Gan15} to value each \ac{VA} contract. 
In each \ac{MC} simulation, even in the calibration stage of the interpolation schemes to value representative 
contracts, we use $10,000$ scenarios. Fewer scenarios results in a noticeable difference, 
as big as $5\%$, between the computed delta value from successive runs. 
In our experiments, we use mortality rates of the 1996 I AM mortality tables provided by the Society of Actuaries.  

We implement the framework in Java and run it on machines with dual quad-core Intel X5355 CPUs. We do not 
use the multiprocessing capability of our machine in these experiments; however, in our future work, 
we will demonstrate that even the serial implementation of our proposed framework can provide 
better efficiency than parallel implementation of \ac{MC} simulations.

\subsection{Representative Contracts}\label{sec:rep_contracts}
As we discuss above in Section \ref{sec:nn}, we do not address the issue of an effective 
sampling method in this paper. Hence, in all of the experiments in this section, we use a simple 
uniform sampling method similar to that in \citep{Hejazi15}. In each set of experiments, we 
select $300$ representative contracts from the set of all \ac{VA} contracts constructed 
from all combinations of points defined in Table \ref{tb:rep_contracts1}. In a set of 
experiments, we select a set of representative contracts at the beginning of the experiment, 
and use the same set for various spatial interpolation methods that we examine in that experiment. 
This allows for a fair comparison between all methods. 

\begin{table}[!bt]
 \centering
 \begin{tabular}{|l|l|}
  \hline
  & Experiment 1\\
  \hline
  Guarantee Type & \{GMDB, GMDB + GMWB\}\\ 
  \hline
  Gender & \{Male, Female\}\\
  \hline
  Age & $\{20, 30, 40, 50, 60\}$\\
  \hline 
  Account Value & $\{1e4, 1e5, 2e5, 3e5, 4e5, 5e5\}$\\
  \hline
  Guarantee Value & $\{0.5e4, 1e5, 2e5, 3e5, 4e5, 5e5, 6e5\}$\\
  \hline 
  Withdrawal Rate & $\{0.04, 0.08\}$\\ 
  \hline 
  Maturity & $\{10, 15, 20, 25\}$\\
  \hline
 \end{tabular}
 \caption{Attribute values from which representative contracts are generated for experiments.}
 \label{tb:rep_contracts1}
\end{table} 

\subsection{Training/Validation Portfolio}
Unlike traditional spatial interpolation schemes, we need to introduce two more portfolios 
to properly train our neural network. In each set of experiments, we select $250$ \ac{VA} contracts 
uniformly at random from the input portfolio as our validation portfolio. 
 
For the training portfolio, we select $200$ contracts uniformly at random from the set of \ac{VA} contracts of all 
combinations of attributes specified in Table \ref{tb:training_portfolio}. The attributes of 
Table \ref{tb:training_portfolio} are intentionally different from the attributes of Table \ref{tb:rep_contracts1} 
to avoid unnecessary overfitting of the data.

\begin{table}[!bt]
 \centering
 \begin{tabular}{|l|l|}
  \hline
  & Experiment 1\\
  \hline
  Guarantee Type & \{GMDB, GMDB + GMWB\}\\ 
  \hline
  Gender & \{Male, Female\}\\
  \hline
  Age & $\{23, 27, 33, 37, 43, 47, 53, 57\}$\\
  \hline 
  Account Value & $\{0.2e5, 1.5e5, 2.5e5, 3.5e5, 4.5e5\}$\\
  \hline
  Guarantee Value & $\{0.5e5, 1.5e5, 2.5e5, 3.5e5, 4.5e5, 5.5e5\}$\\
  \hline 
  Withdrawal Rate & $\{0.05, 0.06, 0.07\}$\\ 
  \hline 
  Maturity & $\{12, 13, 17, 18, 22, 23\}$\\
  \hline
 \end{tabular}
 \caption{Attribute values from which training contracts are generated for experiments.}
 \label{tb:training_portfolio}
\end{table}

\subsection{Parameters Of The Neural Network}\label{sec:nn-param}
In our numerical experiments, we use the following set of parameters to construct and 
train our network. We choose a learning rate of $1$. We set $\mu_{\max}$ in \eqref{eq:momentum-coeff} 
to $0.99$. We use a batch size of $20$ in our training. We fix the seed of the pseudo-random number 
generator that we use to select batches of the training data so that we can reproduce our network 
for a given set of representative contracts, training portfolio, and validation portfolio. Moreover, 
we initialize our weight and bias parameters to zero.

The categorical features in $F^c$ are rider type and 
gender of the policyholder. The following numeric features make up $F^+$.  

\begin{align}\label{eq:nn-numeric-feature}
&f(z, z_i) = \frac{[t(x) - t(x_i)]^{+}}{R_t} \\
&t \in \{\text{maturity, age, AV, GD/AV, GW/AV, withdrawal rate}\} \nonumber
\end{align}
where $AV$ is the account value, $GD$ is the guaranteed death benefit, $GW$ is the guaranteed 
withdrawal benefit, $R_t$ is the range of values that $t$ can assume, $x$ is the vector of numeric attributes for 
input \ac{VA} contract $z$, and $x_i$ is the vector of numeric attributes for representative contract $z_i$. 
The features of $F^-$ are defined in a similar fashion by swapping $x$ and $x_i$ on the right side of 
Equation \eqref{eq:nn-numeric-feature}.

We record \ac{MSE} every $50$ iterations. We compute a moving average with a window of size $10$ to smooth the 
\ac{MSE} values. Moreover, we fit, in a least squares sense, a polynomial of degree $6$ to the smoothed \ac{MSE} 
values and use a window of length $4$ to find the trend in the resulting \ac{MSE} graph. In addition, we choose a 
$\delta$ of $0.005$ as our threshold for the relative error in estimation of the Greeks for the validation 
portfolio.

\subsection{Performance}\label{sec:performance}
In these experiments, we compare the performance (i.e., accuracy, efficiency, and granularity) 
of our proposed neural network scheme, referred to as $NN$ in the results tables, with that of the traditional 
spatial interpolation schemes. From the set of 
interpolation techniques discussed in \citep{Hejazi15}, we choose only the following interpolation 
methods with a corresponding distance function which exhibited the most promising results in \citep{Hejazi15}. 

\begin{itemize}
 \item Kriging with Spherical and Exponential variogram models, 
 \item \ac{IDW} with power parameters $P$ of 1 and 100, 
 \item Gaussian \ac{RBF} with free parameter $\epsilon$ of 1. 
\end{itemize} 
The distance function for the Kriging and \ac{RBF} methods is

\begin{equation*}\label{eq:k_dist}
 D(\mathbf{x}, \mathbf{y}, \gamma) = \sqrt{\sum_{h \in N}\Big(\frac{{x_h - y_h}}{\max_h - \min_h}\Big)^2
 + \gamma \sum_{h \in C}\delta(x_{h}, y_{h})}
\end{equation*}
where $N = \{\text{AV, GD, GW, maturity, age, withdrawal rate}\}$ is the set of numerical values and 
$C = \{ \text{gender, rider} \}$ is the set of categorical values, and $\gamma = 1$.

For the \ac{IDW} methods we choose the following distance function that provided the most promising 
results in \citep{Hejazi15}. 

\begin{align*}\label{eq:r_dist}
&D(\mathbf{x}, \mathbf{y}, \gamma) = \sqrt{f(x_{\text{age}}, y_{\text{age}})g_{\text{age}}(\mathbf{x}, \mathbf{y}) 
+ \sum_{h \in N} g_{\text{h}}(\mathbf{x}, \mathbf{y}) + \gamma \sum_{h \in C}\delta(x_{h}, y_{h})} \nonumber\\
&f(x_{\text{age}}, y_{\text{age}}) = \exp\big(\frac{x_{\text{age}} + y_{\text{age}}}{2} - M\big) \nonumber \\
&g_h(\mathbf{x}, \mathbf{y}) = (\exp(-r_x) x_{h} - \exp(-r_y) y_{h})^2
\end{align*}
where $C = \{\text{gender, rider}\}$, $N = \{\text{maturity, withdrawal rate}\}$, $r = \frac{AV}{GD}$ and 
$M$ is the maximum age in the portfolio.

Because we randomly select our representative contracts according to the method described in 
Section \ref{sec:rep_contracts}, we compare the performance of the interpolation schemes using 
$6$ different realizations of the representative contracts, $Si, 1 \leq i \leq 6$. For our proposed 
neural network approach, we use the same training portfolio and validation portfolio in all of 
these $6$ experiments. We study the effect of the training portfolio and the validation portfolio 
in a different experiment.

Table \ref{tb:rel_err1} displays the accuracy of each scheme in estimation of the delta value for the 
input portfolio. The accuracy of different methods is recorded as the relative error 

\begin{equation}\label{eq:relative_err}
 \text{Err}_m = \frac{\Delta_m - \Delta_{MC}}{|\Delta_{MC}|}
\end{equation}
where $\Delta_{MC}$ is the estimated delta value of the input portfolio computed by \ac{MC} simulations 
and $\Delta_m$ is the estimate delta value of the input portfolio computed by method $m$. The results of Table 
\ref{tb:rel_err1} show the superior performance of our proposed neural network (NN) framework in terms of accuracy. 
Except in a few cases, the accuracy of our proposed NN framework is better than all of the other interpolation schemes. 
The Spherical Kriging has the best performance amongst the traditional interpolation schemes. Comparing the 
accuracy results of our proposed neural network scheme with that of Spherical Kriging shows that 
the relative error of the proposed scheme has lower standard deviation and hence is more reliable.

\begin{table}[!bt]
 \centering
 \begin{tabular}{|l|l|l|l|l|l|l|}
  \hline
  \multirow{2}{*}{Method} & \multicolumn{6}{|c|}{Relative Error (\%)}\\
  \cline{2-7} 
  & S1 & S2 & S3 & S4 & S5 & S6\\
  \hline
  Kriging (Sph) & $-0.60$ & $0.55$ & $7.62$ & $2.86$ & $2.58$ & $3.59$\\ 
  \hline
  Kriging (Exp) & $-0.29$ & $1.73$ & $8.09$ & $4.77$ & $3.46$ & $4.38$ \\
  \hline
  IDW (p = 1) & $-21.43$ & $-14.48$ & $-21.76$ & $-7.22$ & $-12.47$ & $-11.77$ \\
  \hline
  IDW (p = 100) & $-11.74$ & $-7.81$ & $-4.36$ & $-0.07$ & $-2.72$ & $-2.45$ \\ 
  \hline 
  RBF (Gau, $\epsilon = 1$) & $-0.76$ & $-5.21$ & $-10.45$ & $-7.83$ & $2.47$ & $4.11$ \\
  \hline
  NN & $-1.16$ & $0.84$ & $1.56$ & $0.85$ & $1.66$ & $-1.46$\\
  \hline
 \end{tabular}
 \caption{Relative error in estimation of the portfolio's delta value by each method.}
 \label{tb:rel_err1}
\end{table}

In Table \ref{tb:sim_time1}, the average training and estimation time of each method is presented for two 
scenarios: (1) the method is used to estimate only the delta value of the entire portfolio and (2) the 
method is used to estimate the delta value of each policy in the input portfolio.
Because of the complex calculations required to train the proposed NN method, the running time of 
the proposed NN method is longer than that of the traditional interpolation scheme. However, it still outperforms 
the \ac{MC} simulations (speed up of $> \times 15$). 

In this experiment, assuming no prior knowledge of the market, we used the value of zero as our initial value for weight/bias 
parameters which is far from the optimal value and causes the performance of the proposed NN method to suffer from a long 
training time. In practice, insurance companies estimate the Greeks of their portfolios on frequent intraday basis to do dynamic 
hedging. Assuming a small change in the market condition, one does not expect the Greek values of the \ac{VA} policies 
to change significantly. Hence, intuitively, the change in the 
optimal values of weight/bias parameters of the network under the previous and the current market conditions should be 
small. In our future paper, in the context of estimating the probability distribution of the one year loss, 
we will demonstrate how we exploit this fact to reduce the training time of the network from an average of $4000$ iterations 
to less than $200$ iterations  and hence reduce the training time significantly. In particular, assuming a neural network that has been 
calibrated to the previous market conditions, we construct a new network that uses the values of the weight/bias parameters 
of the previous network as the initial values for the weight/bias parameters in the training stage.

\begin{table}[!bt]
 \centering
 \begin{tabular}{|l|l|l|}
  \hline
  Method & Portfolio & Per Policy\\
  \hline
  MC & $10617$ & $10617$\\
  \hline
  Kriging (Spherical) & $41$ & $\gg 10617$\\ 
  \hline
  Kriging (Exponential) & $41$ & $\gg 10617$ \\
  \hline
  IDW (P = 1) & $29$ & $29$\\
  \hline
  IDW (P = 100) & $28$ & $28$\\ 
  \hline 
  RBF (Gaussian, $\epsilon = 1$) & $41$ & $41$\\
  \hline
  NN & $539$ & $539$ \\
  \hline
 \end{tabular}
 \caption{Simulation time of each method to estimate the delta value. All times are in seconds.}
 \label{tb:sim_time1}
\end{table}

\begin{figure}[bt]
\centering
\includegraphics[width=\textwidth, height=0.4\textheight]{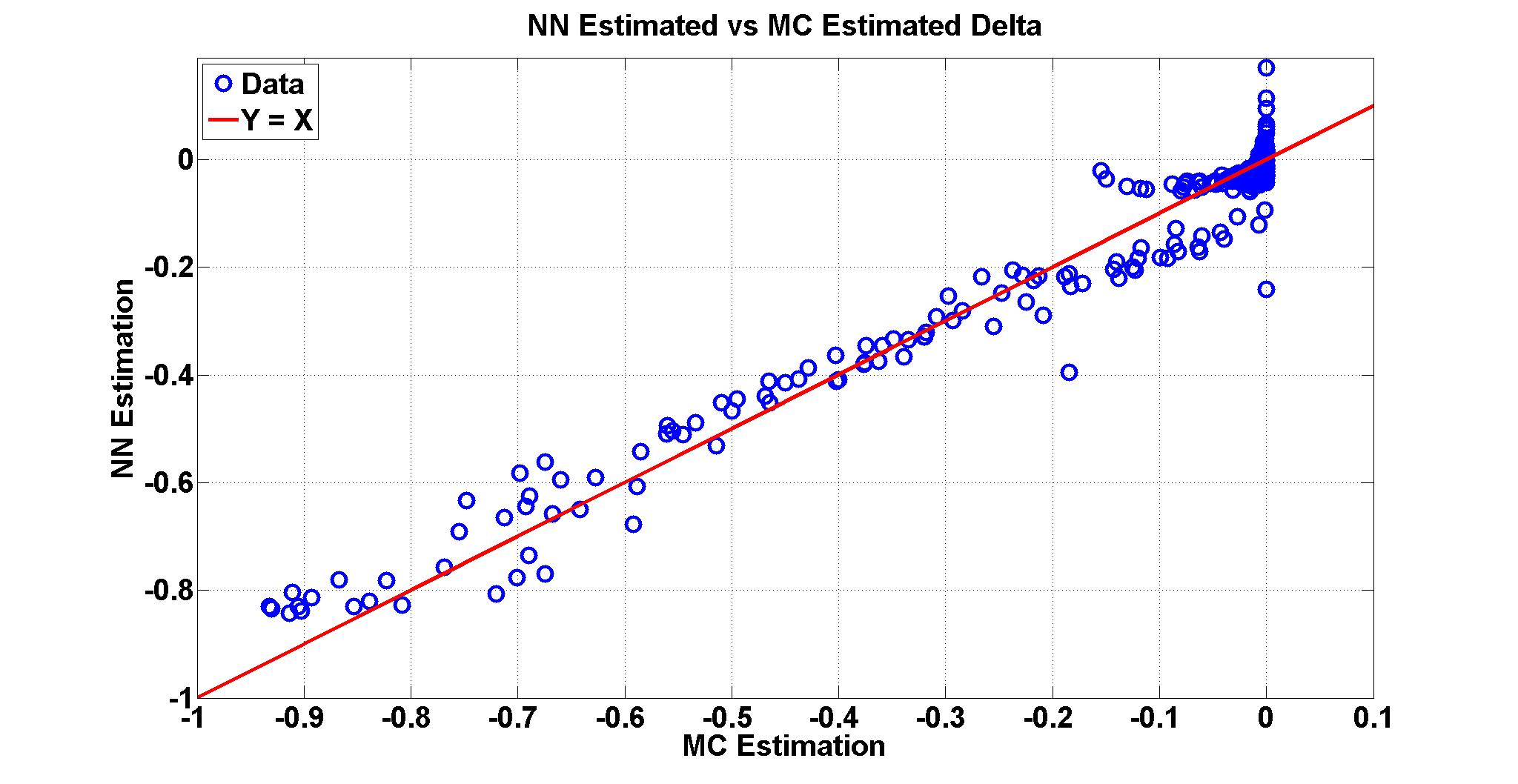}
\caption{Estimated delta values of the validation portfolio by \ac{MC} simulations and the neural network.}
\label{fig:per-policy-err}
\end{figure}

A comparison of the running time in the two columns of 
Table \ref{tb:sim_time1} shows that the proposed NN method, similar to \ac{IDW} and \ac{RBF}, can 
be used to efficiently provide a granular view of the delta values in the input portfolio. Figure 
\ref{fig:per-policy-err} shows a granular view of the estimated delta values by our proposed NN scheme for 
the validation portfolio. As shown in the figure, the NN estimated delta values closely follow their 
corresponding \ac{MC} estimated values (plotted data values are very close to the line $y = x$). 
In particular, the majority of data points are within a distance of $0.1$ of the $y = x$ line (the red line). 
Moreover, the data points are distributed almost uniformly around the red line. In other words, the amount 
of over estimations by the neural network is close to the amount of under estimations by the neural network. 
Therefore, the estimation errors, in aggregate, cancel each other out, resulting in a smaller portfolio error 
than might be expected from looking at the maximum absolute error alone in Figure \ref{fig:per-policy-err}. 

The graph also shows a cluster of points around the origin. In fact, 
the majority of the points are very close to the origin with only a few points deviating relatively far in 
each direction. These few points do not significantly affect the accuracy, especially when the estimations 
are accurate for big delta values. 

\subsection{Sensitivity to Training/Validation Portfolio}\label{sec:ne-t-v-sensitivity}
The training of our proposed NN method requires the selection of three \ac{VA} portfolios. In the 
experiments of Section \ref{sec:performance}, we fix the selection of two of these portfolios 
(i.e., training portfolio and validation portfolio) while we measured the performance of our 
proposed method by changing the set of representative contracts. In the experiments of this section, 
we investigate the sensitivity of the accuracy and efficiency of our proposed method to the choice of 
training and validation portfolio. We conduct two sets of experiments in which we fix the choice of 
representative contracts and either the training portfolio or the validation portfolio, while training the network 
with different realizations of the remaining portfolio. In the first set of experiments, we 
fix the choice of the representative contracts and the validation portfolio. We train the network 
with $5$ different choices of the training portfolio and estimate the delta value of the input portfolio 
in each experiment. In the second set of experiments, we fix the choice of the representative contracts and 
the training portfolio and train the network with $5$ different realizations of the validation portfolio. We 
then use the trained network to estimate the delta value of the input portfolio in each experiment. We used the 
same set of representative contracts in both set of experiments. 

\begin{table}[!bt]
 \centering
 \begin{tabular}{|l|l|l|l|l|}
  \hline
  \multirow{2}{*}{Variable Portfolio} & \multicolumn{2}{|c|}{Relative Error (\%)} & \multicolumn{2}{|c|}{Running Time}\\
  \cline{2-5} 
  & Mean & STD & Mean & STD\\
  \hline 
  Training & $0.27$ & $1.52$ & $660$ & $246$\\
  \hline
  Validation & $-0.62$ & $1.51$ & $523$ & $38$\\
  \hline
 \end{tabular}
 \caption{Statistics on the running time sensitivity and accuracy sensitivity of the training network with different sets of training and 
validation portfolios. The recorded errors are relative errors as defined in \eqref{eq:relative_err}. All times are in seconds.}
 \label{tb:sim_train_valid_stat}
\end{table}

The statistics for the running time (training and estimation) and accuracy of each set of experiments are presented in Table 
\ref{tb:sim_train_valid_stat}. The relatively big values of the standard deviations indicate that the accuracy of the estimation 
is sensitive to the choice of the training portfolio and the validation portfolio. Despite this sensitivity, the method 
remains accurate. 

The choice of the training portfolio can significantly affect the running time of the neural network; however, the running 
time of the network is fairly insensitive to changes in the choice of validation portfolio. The validation portfolio in the training 
is mainly used as a guard against over fitting. It is a useful stopping criteria to fine tune the network once we are  
close to the local optimum. But the training portfolio controls the path that the training takes to arrive at a 
neighborhood close to the local optimum. A bad training portfolio can slow the training by introducing a large deviation from a consistent 
path towards the local optimum. Hence the choice of the training portfolio has a more significant effect than the choice of 
the validation portfolio on the running time. 

\subsection{Sensitivity to Sample Sizes}\label{sec:sample_size}
In the previous experiments, we examined the sensitivity of our proposed neural network framework to the selection of 
the training portfolio, the validation portfolio, and the set of representative contracts. In this section, 
we conduct experiments that assess the sensitivity of our proposed framework on the size of these portfolios. In each 
experiment, we fix two out of the three required portfolios while changing the size of the third portfolio. 
For each selected size of the latter portfolio, we train the network with $5$ realizations of the portfolio and record 
the running time and accuracy of the method. 

Table \ref{tb:portfolio-size-stat} contains the statistics on the recorded running time and the relative error for each set of 
selected portfolio sizes. Each row in the table begins with a tuple $(r, t, v)$ denoting the size of the set of representative 
contracts, the training portfolio, and the validation portfolio, respectively. In the scenarios corresponding to the first three rows, 
we changed the size of representative contracts. The second three rows show the results for the scenarios in which we changed the 
size of the training portfolio. Finally, in the scenarios corresponding to the third three rows, we changed the size of the validation portfolio.

\begin{table}[!bt]
 \centering
 \begin{tabular}{|l|l|l|l|l|}
  \hline
  \multirow{2}{*}{Portfolio Sizes} & \multicolumn{2}{|c|}{Relative Error (\%)} & \multicolumn{2}{|c|}{Running Time}\\
  \cline{2-5} 
  & Mean & STD & Mean & STD\\
  \hline 
  $(300, 200, 250)$ & $0.38$ & $1.35$ & $539$ & $120$\\
  \hline
  $(250, 200, 250)$ & $-0.73$ & $1.42$ & $373$ & $73$\\
  \hline
  $(200, 200, 250)$ & $-1.62$ & $1.52$ & $310$ & $85$\\
  \hline
  \hline
  $(300, 200, 250)$ & $0.27$ & $1.52$ & $539$ & $246$\\
  \hline
  $(300, 150, 250)$ & $-4.31$ & $7.66$ & $708$ & $254$\\
  \hline
  $(300, 100, 250)$ & $6.50$ & $14.47$ & $669$ & $303$\\
  \hline
  \hline
  $(300, 200, 250)$ & $-0.62$ & $1.51$ & $523$ & $38$\\
  \hline
  $(300, 200, 200)$ & $0.70$ & $3.23$ & $511$ & $24$\\
  \hline
  $(300, 200, 150)$ & $2.31$ & $3.67$ & $582$ & $188$\\
  \hline
 \end{tabular}
 \caption{Statistics on running time sensitivity and accuracy sensitivity of training network with portfolios of various sizes. 
The recorded errors are relative errors as defined in \eqref{eq:relative_err}. All times are in seconds.}
 \label{tb:portfolio-size-stat}
\end{table}

The results of Table \ref{tb:portfolio-size-stat} show that decreasing the number of representative contracts increases the 
efficiency of the network. Furthermore, the amount of decrease in running time is proportional to the amount of decrease in 
the number of representative contracts. This result is expected since the number of calculations in the network is proportional 
to the number of neurons in the hidden layer which is proportional to the number of representative contracts. Although the accuracy 
of the method deteriorates as we decrease the number of representative contracts, the accuracy of the worst network is still comparable 
to the best of the traditional spatial interpolation techniques (see Table \ref{tb:rel_err1}). Hence, if required, we can sacrifice 
some accuracy for better efficiency. 

According to the statistics in the second set of three rows of Table \ref{tb:portfolio-size-stat}, decreasing the size of the training portfolio can 
significantly affect the accuracy of the method. Decreasing the number of training \ac{VA} contracts results in a poorer coverage of the space 
in which the network is trained. In the space where the training \ac{VA} contracts are sparse, the parameters of the representative 
\ac{VA} contracts are not calibrated well, resulting in poor accuracy of estimation. 
Although the mean of the simulation time does not consistently decrease with the decrease in the size of the training portfolio, the standard 
deviation of the simulation time increases significantly. The increase in the standard deviation of the network's simulation time 
is a further proof that the network is struggling to calibrate its parameters for the smaller training portfolios. 

The results in the last three rows of Table \ref{tb:portfolio-size-stat} suggest that decreasing the size of validation portfolio decreases 
the accuracy of the proposed framework. The deterioration in the performance of the network is more apparent from the amount of increase 
in the standard deviation of the relative error values. As one decreases the size of the validation portfolio, the \acp{VA} in the 
validation portfolio provide a poorer representation of the input portfolio. Although the change in the accuracy of the method is significant, 
the running time of the method is less affected by the size of the validation portfolio, except for the validation portfolio of the smallest 
size, where one can see a big increase in the standard deviation of the running time. 

As we mentioned earlier in Section \ref{sec:ne-t-v-sensitivity}, the validation portfolio only affects the last stage of the 
training where the network parameters are close to their local optimal values. When 
the size of the validation portfolio is too small, various realizations of the validation portfolio may not adequately fill the space resulting 
in portfolio delta values that differ significantly from one realization to another. Hence, the overlap between the $\delta$ neighborhood 
of the portfolio delta values for various validation portfolios and the local neighborhood of the optimal network parameter values may vary 
in place and size significantly. The network stops the training as soon as it finds a set of network parameters that are within the 
aforementioned common neighborhood. Therefore, the training time of the network can vary significantly based on the size and the place of the 
common neighborhood. As the size of the common neighborhood increases, the network spends less time searching for a set of network parameters that 
are within the common neighborhood. Because the training time is a significant part of the running time of the proposed neural network scheme, 
the standard deviation of the running time increases as the result of the increase in the standard deviation of the training time.  

%% file: conclusion.tex
\section{Concluding Remarks}\label{sec:conclusion}
In recent years, a spatial interpolation framework has been proposed to improve the 
efficiency of valuing large portfolios of complex insurance products, 
such as \ac{VA} contracts via nested \ac{MC} simulations \citep{Gan13-2, Gan15, Hejazi15}. 
In the proposed framework, a small set of representative \ac{VA} contracts is selected and 
valued via \ac{MC} simulations. The values of the representative contracts 
are then used in a spatial interpolation method that finds the value of the 
contracts in the input portfolio as a linear combination of the values of the 
representative contracts. 

Our study of traditional spatial interpolation techniques (i.e., Kriging, \ac{IDW}, 
\ac{RBF}) \citep{Hejazi15} highlights the strong dependence of the accuracy of 
the framework on the choice of distance function used in the estimations. Moreover, 
none of the traditional spatial interpolation techniques can provide us with 
all of accuracy, efficiency, and granularity, as defined in \citep{Hejazi15}. 

In this paper, we propose a neural network implementation of the spatial interpolation 
technique that learns an effective choice of the distance function and provides 
accuracy, efficiency, and granularity. We study the performance of the proposed 
approach on a synthetic portfolio of \ac{VA} contracts with GMDB and GMWB riders. 
Our results in Section \ref{sec:ne} illustrate the superior accuracy of our 
proposed neural network approach in estimation of the delta value for the input 
portfolio compared to the traditional spatial interpolation techniques. 

Although the proposed NN framework, compared with traditional spatial interpolation 
techniques, requires longer training time, it can interpolate the delta values of 
an input portfolio of size $N$ in a time proportional to  $n \times N$ ($n$ is 
the number of samples), which is essentially the same as the time required by the 
most efficient traditional spatial interpolation techniques \ac{IDW} and \ac{RBF}. 
Moreover, the proposed NN approach provides an efficient 
solution to the issue of choosing a distance function. 

Training of the network requires us to introduce two additional sets of sample \ac{VA} contracts, 
i.e., the training portfolio and the validation portfolio, compared to the traditional 
frameworks \citep{Hejazi15}. Our experiments in Section \ref{sec:ne} show that, 
if each of the aforementioned sample sets is sufficiently large, a random selection of 
these sets from a predefined set of \ac{VA} contracts that uniformly covers the space of 
the input portfolio does not significantly affect the accuracy and efficiency of the method. However 
the size of each of these sample sets can significantly affect the performance of our proposed 
neural network approach. 

Although this paper studies an important application of the proposed neural network framework, in the 
future, we will extend the approach to compute other key risk metrics. In particular, we will demonstrate 
the efficiency and accuracy of this framework in estimating the probability distribution of the portfolio loss, 
which is key to calculate the Solvency Capital Requirement. 

%% file: ack.tex
\section{Acknowledgements}
This research was supported in part by the Natural Sciences and Engineering Research Council of Canada (NSERC).

%% file: appendix.tex
\section{How To Choose The Training Parameters}\label{sec:parameters}
The training method that we discuss in Section \ref{sec:nn} is dependent on the choice of numerous free parameters such as 
the learning rate and $\mu_{\max}$. In this appendix, we discuss heuristic ways to choose a value for each of these free parameters 
and justify each choice. 

In order to determine a good choice of the learning rate and the batch size, we need to 
train the network for some number of iterations, say $3000$, and study the error associated with the training portfolio as a 
function of the number of iterations. If the graph has a general decreasing trend and it does not have many big jumps between 
consecutive iterations, then we say that the choice of the learning rate/batch size is stable. Otherwise, we call the choice 
of the learning rate/batch size unstable. 

\begin{figure}[!bt]
 \centering
 \begin{subfigure}{0.47\textwidth}
  \includegraphics[trim=3cm 1cm 5cm 0cm, clip=true, width=\linewidth, height=0.3\textheight]{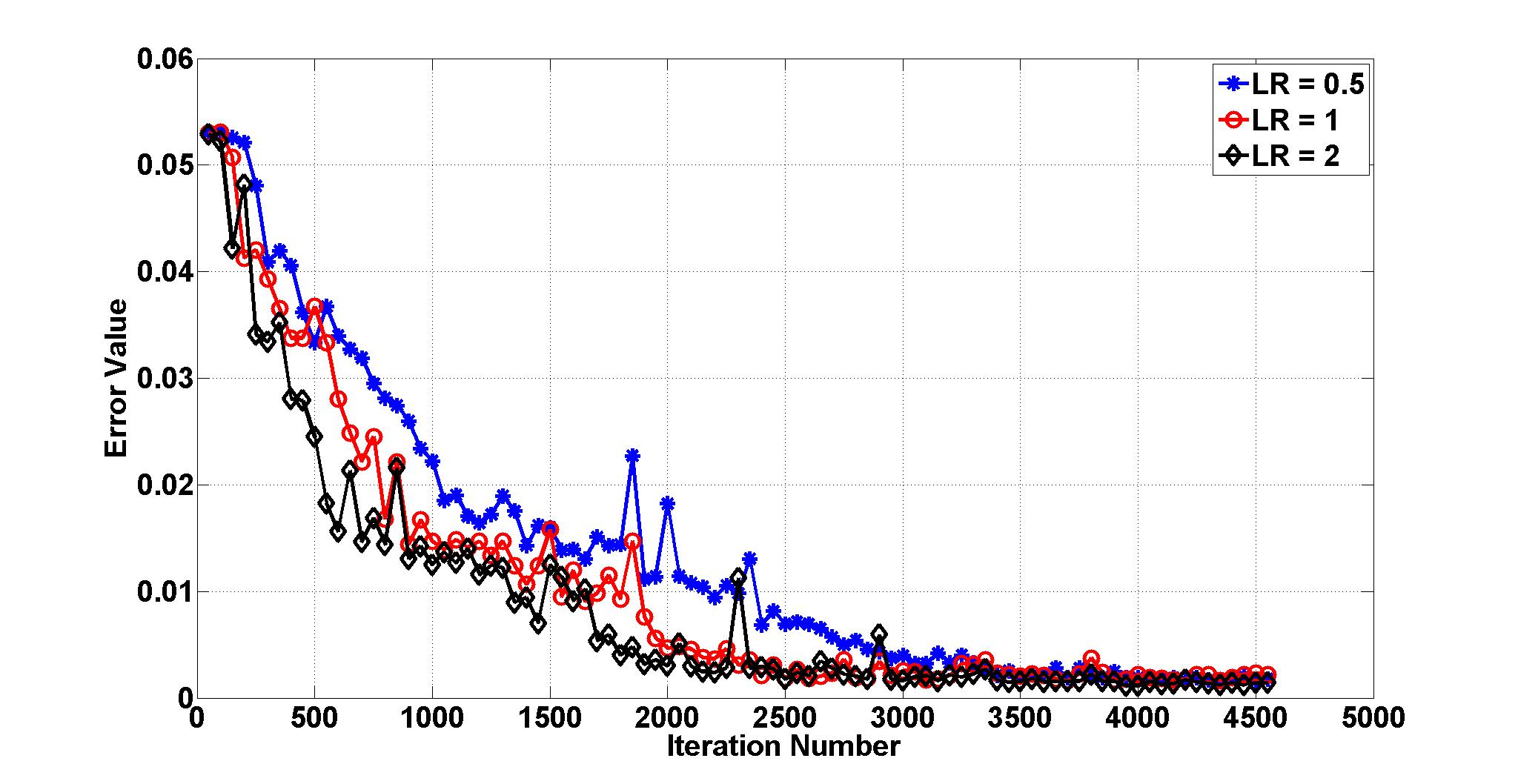}  
 \end{subfigure}
 \begin{subfigure}{0.47\textwidth}
  \includegraphics[trim=3cm 0.5cm 5cm 0cm, clip=true, width=\linewidth, height=0.3\textheight]{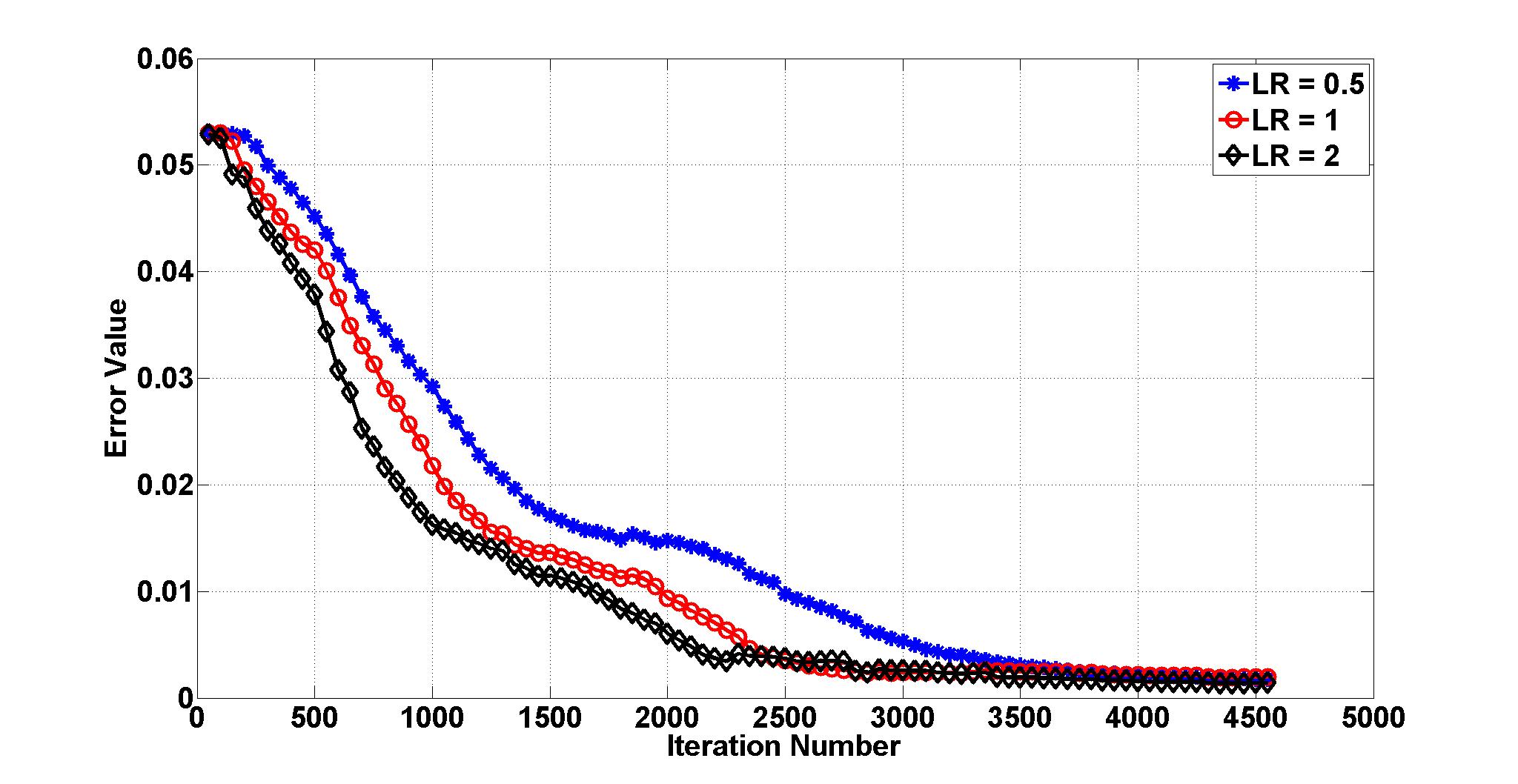}  
 \end{subfigure}
 \caption{The MSE error graph (left) and the moving average smoothed MSE error graph (right) of the training portfolio as a function of iteration number and learning rate.}
 \label{fig:error-learn-rate}
\end{figure}

From \eqref{eq:gradient-descent}, we see that the choice of the learning rate parameter affects the amount of change in 
the weight and bias parameters per iteration. As we discuss in Section \ref{sec:net-train}, too small of a change 
increases the training time while too big of a change causes numerical instability. From Figure \ref{fig:error-learn-rate}, 
we see that, as we increase the value of the learning rate, the graph of the error values moves downwards which means that 
the training has sped up. However, for a learning rate equal to $2$, we see many big jumps in the graph which suggests 
numerical instability. The numerical instability is more obvious from the moving average smoothed curve of error values. 
More specifically, starting from iteration $2000$, the smoothed MSE error graph for a learning rate of $2$ has big jumps 
which are signs of numerical instability. Note that the smoothed MSE error graphs for learning rates $0.5$ and $1$ are 
much smoother. 

Although the \ac{MSE} error graphs for learning rates 0.5 and 1 are stable, these graphs still contain big jumps, which are the 
result of using mini-batch training. As we discuss earlier, at each iteration, the mini-batch training method updates the weights in the 
direction of the gradient for the error function \eqref{eq:mini-batch-error}. As the definition suggests, the error \eqref{eq:mini-batch-error} 
is dependent on the choice of the VA contracts in that particular mini-batch. Therefore, the direction of the gradient for this function may not align 
with the gradient of \eqref{eq:mse-error} which is defined over the complete batch (i.e., the training portfolio) and hence updating the weights 
along the direction of the gradient of \eqref{eq:mini-batch-error} can move the weights away from their local minimum which contributes to the jumps 
that we see in the MSE error graph. Too many such misalignments can cause big jumps in the MSE error graph. 
Introducing momentum can to some extent ameliorate this phenomenon but it cannot completely avoid it \citep{Murphy12, Sutskever13}. 
Furthermore, as we discuss in detail below, the size of the mini-batch that we choose affects the severity of these jumps. Hence, 
one should expect some jumps in the MSE error graph which determines the MSE error for estimation of the key risk metric of interest for all the VAs 
in the training portfolio. However, if big jumps happen too often, they may cause the weights to oscillate around the 
local minimum of the MSE error function or, even worse, they may force the weights to move away from the local minimum. 
Choosing an appropriate size for the mini-batch (as described below) usually helps to ameliorate this potential deficiency.

To find a good choice of the learning rate, we can start from a value of $1$ for the learning rate and determine if that choice is stable? 
If the choice of learning rate is stable, we double the value of the learning rate and repeat the process until we find a learning rate which 
is unstable. At this point, we stop and choose the last stable value of the learning rate as our final choice of the learning rate. If 
the learning rate equal to $1$ is unstable, we decrease the value of learning rate to half of its current value and repeat this 
process until we find a stable learning rate. 

The batch size controls the speed of training and the amount of error that we get in approximating the gradient of the MSE for the 
entire training set. Small batch sizes increase the speed of training; however, they also increase the amount of error in approximating the 
gradient of the MSE error. A good batch size should be small enough to increase the speed of training but not so small as to introduce a big 
approximation error. To find a good batch size value, we start with a small value, say $5$, and determine if this choice of batch size 
is stable. If so, we stop and choose it as our final choice of the batch size. If the batch size is unstable, 
we double the batch size and repeat the process until we find a stable batch size. 

\begin{figure}[!bt]
 \centering
 \begin{subfigure}{0.47\textwidth}
  \includegraphics[trim=3cm 1cm 5cm 0cm, clip=true, width=\linewidth, height=0.3\textheight]{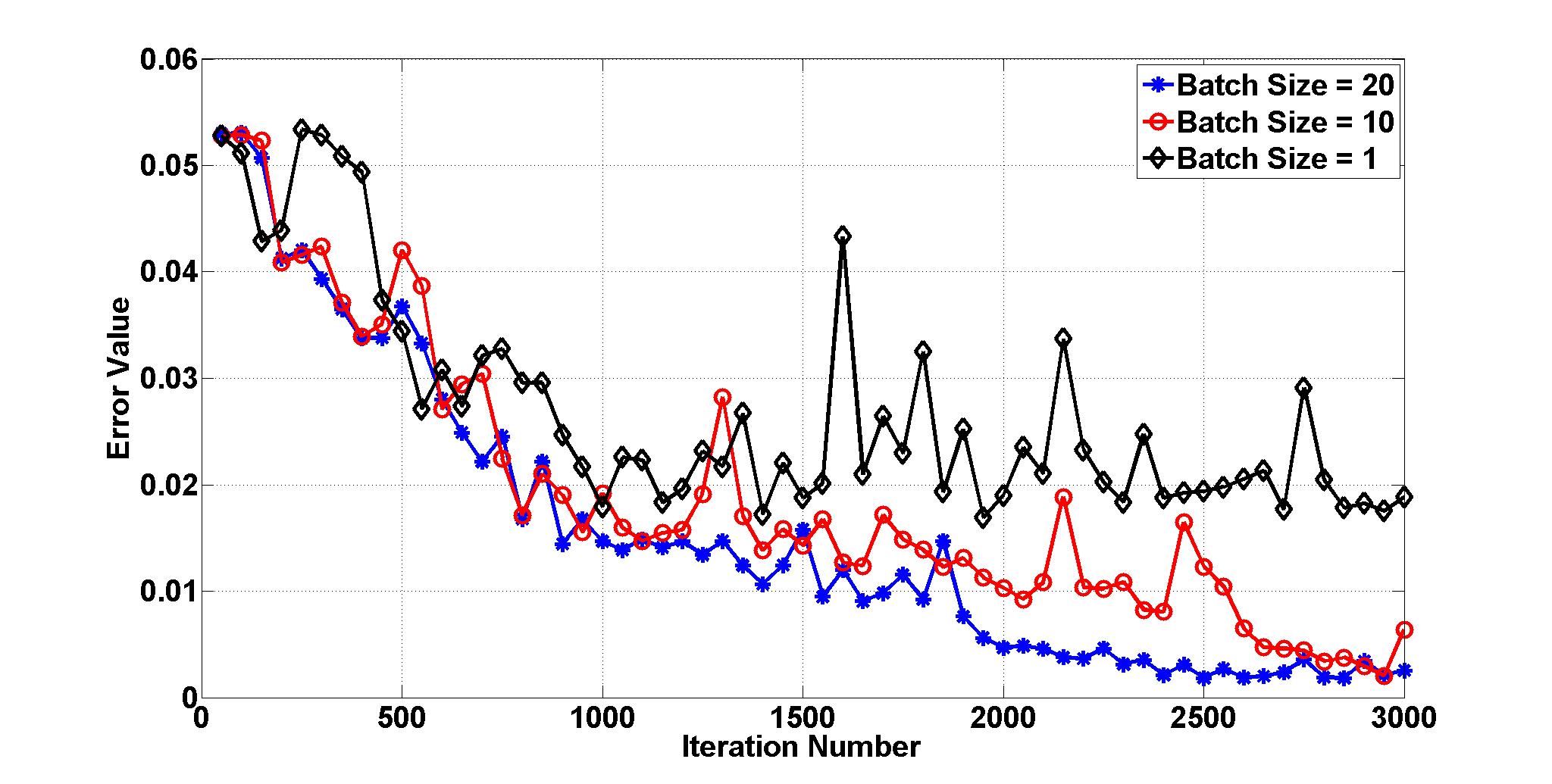}  
 \end{subfigure}
 \begin{subfigure}{0.47\textwidth}
  \includegraphics[trim=3cm 0.5cm 5cm 0cm, clip=true, width=\linewidth, height=0.3\textheight]{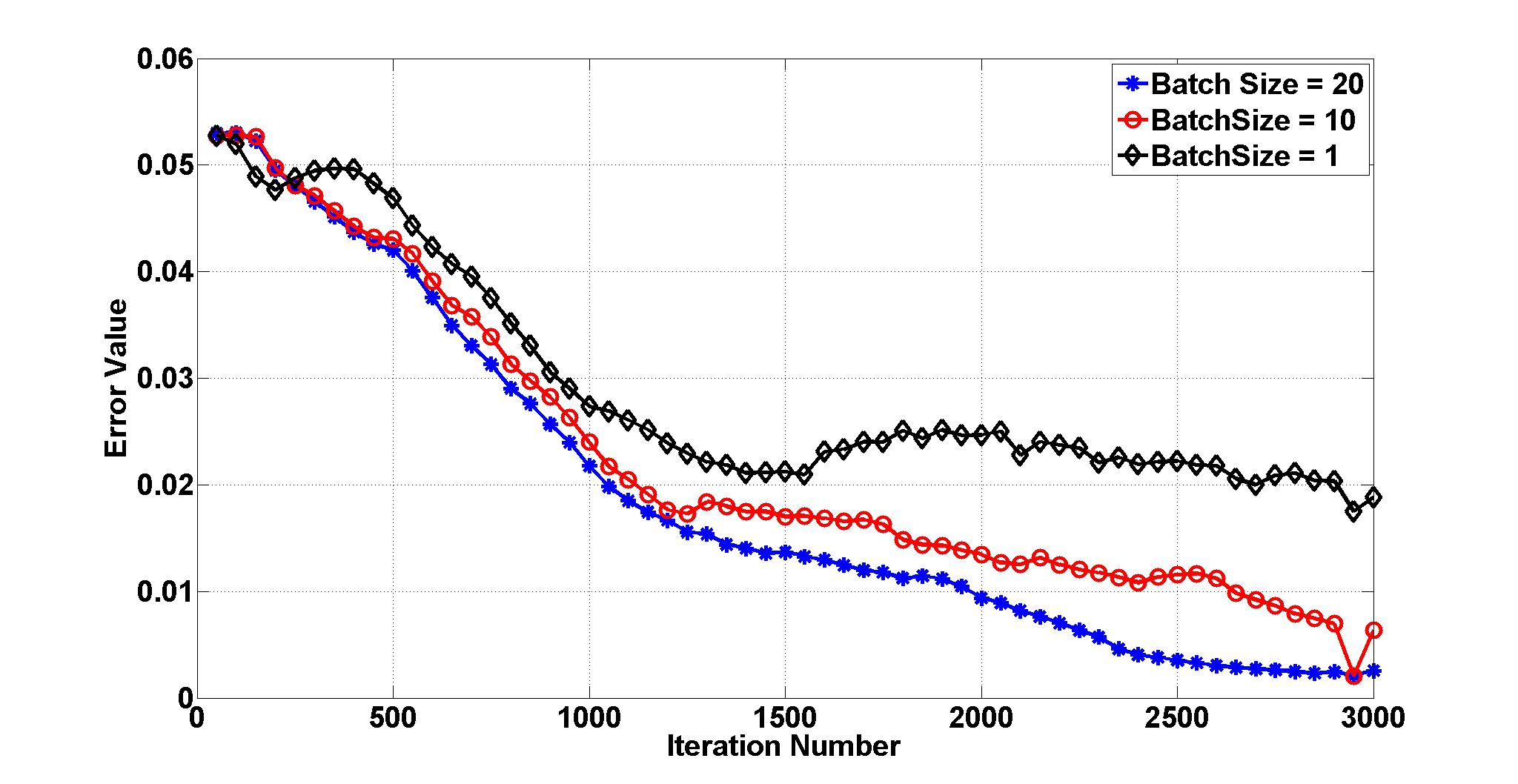}  
 \end{subfigure}
 \caption{The MSE error graph (left) and the moving average smoothed MSE error graph (right) of the training portfolio as a function of iteration number and batch-size.}
 \label{fig:error-batch-size}
\end{figure}

Figure \ref{fig:error-batch-size} shows that small batch size values are associated with many big jumps and hence are unstable. As we increase the 
batch size value, the error graph becomes much more stable-- fewer jumps and a more consistent decreasing trend. 

Notice that, in the aforementioned processes for finding the appropriate value of the learning rate and batch size, doubling the values may seem too aggressive 
as the values may increase or decrease too quickly. To alleviate this problem, upon reaching a desired value, we can do a binary search 
between the final choice of parameter's value and the next best choice (the value of parameter before the final doubling) of the parameter's value. 

Nesterov, \citep{Nesterov03, Nesterov83}, advocates a constant momentum coefficient for strongly convex functions and advocates Equation 
\eqref{eq:nesterov-momentum-coeff} when the function is not strongly convex \citep{Sutskever13}. 

\begin{equation}\label{eq:nesterov-momentum-coeff}
\mu_t = 1 - \frac{3}{t + 5}
\end{equation}

Equation \eqref{eq:momentum-coeff}, 
suggested in \citep{Sutskever13}, blends a proposal similar to  Equations \eqref{eq:nesterov-momentum-coeff} and a constant momentum coefficient. 
Equation \eqref{eq:nesterov-momentum-coeff} converges quickly to values very close to $1$. In particular, for $t \geq 25$, $\mu_t \geq 0.9$. 
Hence, as suggested in \citep{Sutskever13}, we should choose large values ($0.9, 0.99, 0.995, 0.999$)  of $\mu_{\max}$ to achieve better performance 
and that is what we suggest too. 

\begin{figure}[!bt]
 \centering
 \begin{subfigure}{0.47\textwidth}
  \includegraphics[trim=3cm 1cm 5cm 0cm, clip=true, width=\linewidth, height=0.3\textheight]{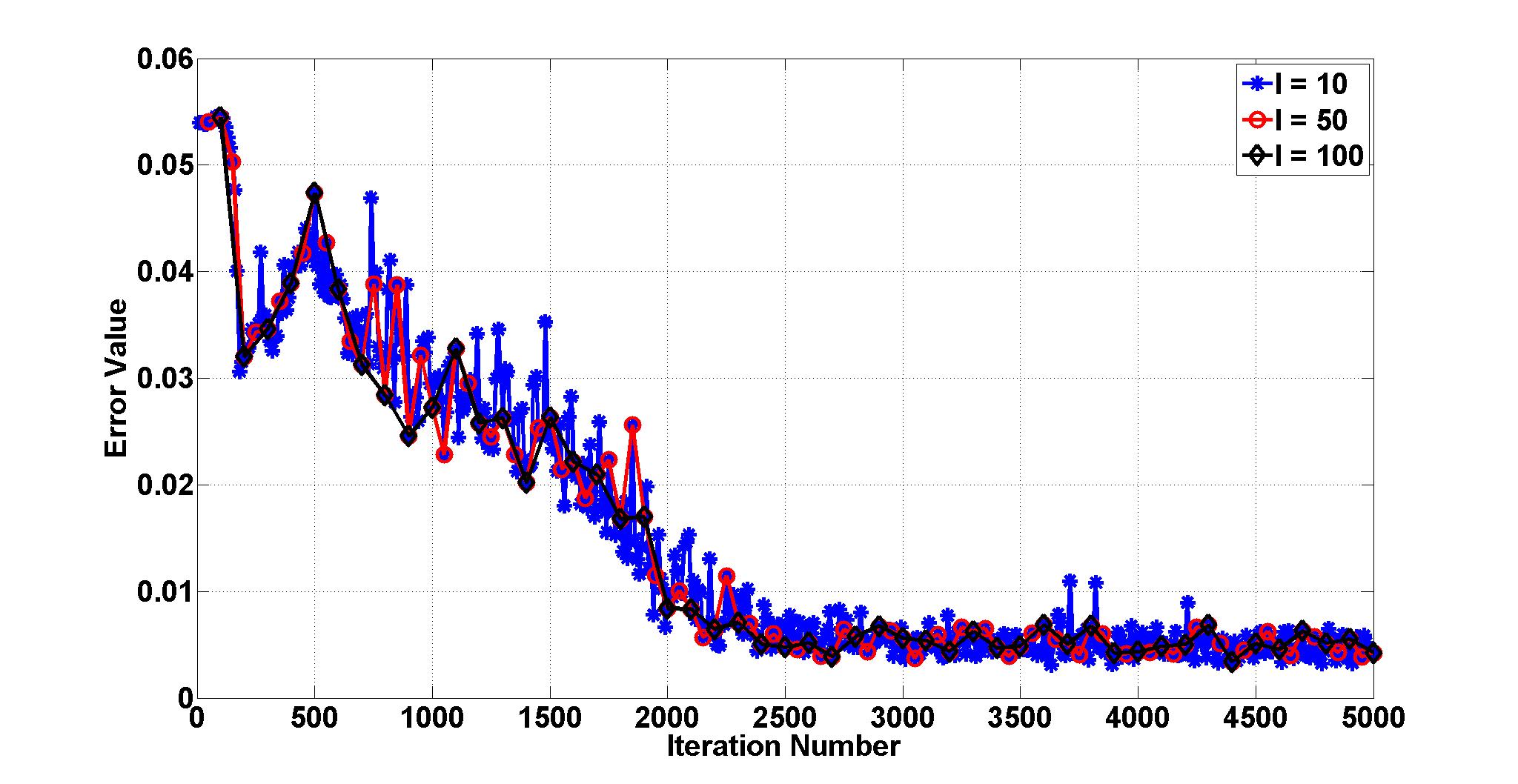}  
 \end{subfigure}
 \begin{subfigure}{0.47\textwidth}
  \includegraphics[trim=3cm 0.5cm 5cm 0cm, clip=true, width=\linewidth, height=0.3\textheight]{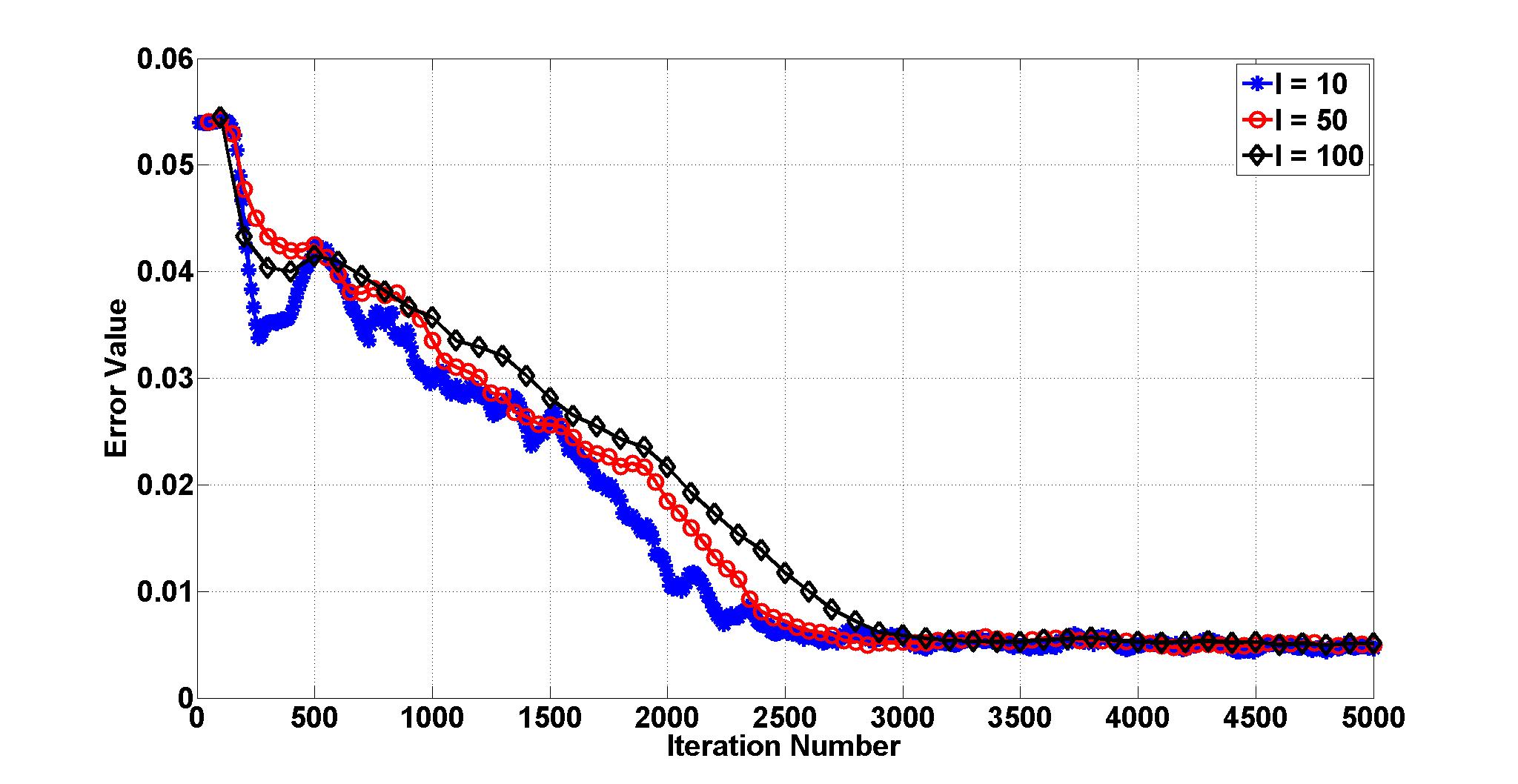}  
 \end{subfigure}
 \caption{The MSE error graph (left) and the moving average smoothed MSE error graph (right) of the validation portfolio as a function of iteration number and $I$ value.}
 \label{fig:error-I}
\end{figure}

In Section \ref{sec:stop-cond}, we proposed a mechansim to detect stopping events and avoid over-training of the network. As part of this mechanism, we need 
to record the MSE of the validation set every $I^{th}$ iteration. Too small values of $I$ can slow down the training process while too big values of $I$ can 
result in losing the information regarding the trend that exists in the MSE graph. In order to find a good value of $I$ that neither slows down the training 
too much nor creates excessive information loss, we can use a multiplicative increase process similar to that described above for the batch size. We start with 
a small value of $I$, say $10$, and train the network for some $4000$ iterations and draw the graph of MSE values. We then double the $I$ value and graph the 
MSE for the new value of $I$. If the MSE graph for the new value of $I$ has a similar trend as the MSE graph for the previous value of $I$, we keep increasing 
the $I$ value and repeat the process. But, if the resulting graph has lost significant information regarding increasing/decreasing trends in the previous 
graph, then we stop and choose the previous value of $I$ as the appropriate choice of $I$. For example, in Figure \ref{fig:error-I}, the MSE graph corresponding 
to the value of $100$ has fewer big valleys and big peaks than the MSE graph for the value of $50$. Hence, we have lost a significant amount of information regarding 
trends in the graph. However, the MSE graph for the value of $10$ has a roughly similar number of big valleys and big peaks compared with the MSE graph 
for the value of $50$. Hence, the value of $50$ is a much better choice for $I$ than either $100$ or $10$. The value of $50$ allows for a faster training 
than the value of $10$ and has more information regarding increasing/decreasing trends in the MSE error graph than the value of $100$. 

\begin{figure}[!bt]
 \centering
 \begin{subfigure}{0.70\textwidth}
  \includegraphics[trim=3cm 1cm 5cm 0cm, clip=true, width=\linewidth, height=0.3\textheight]{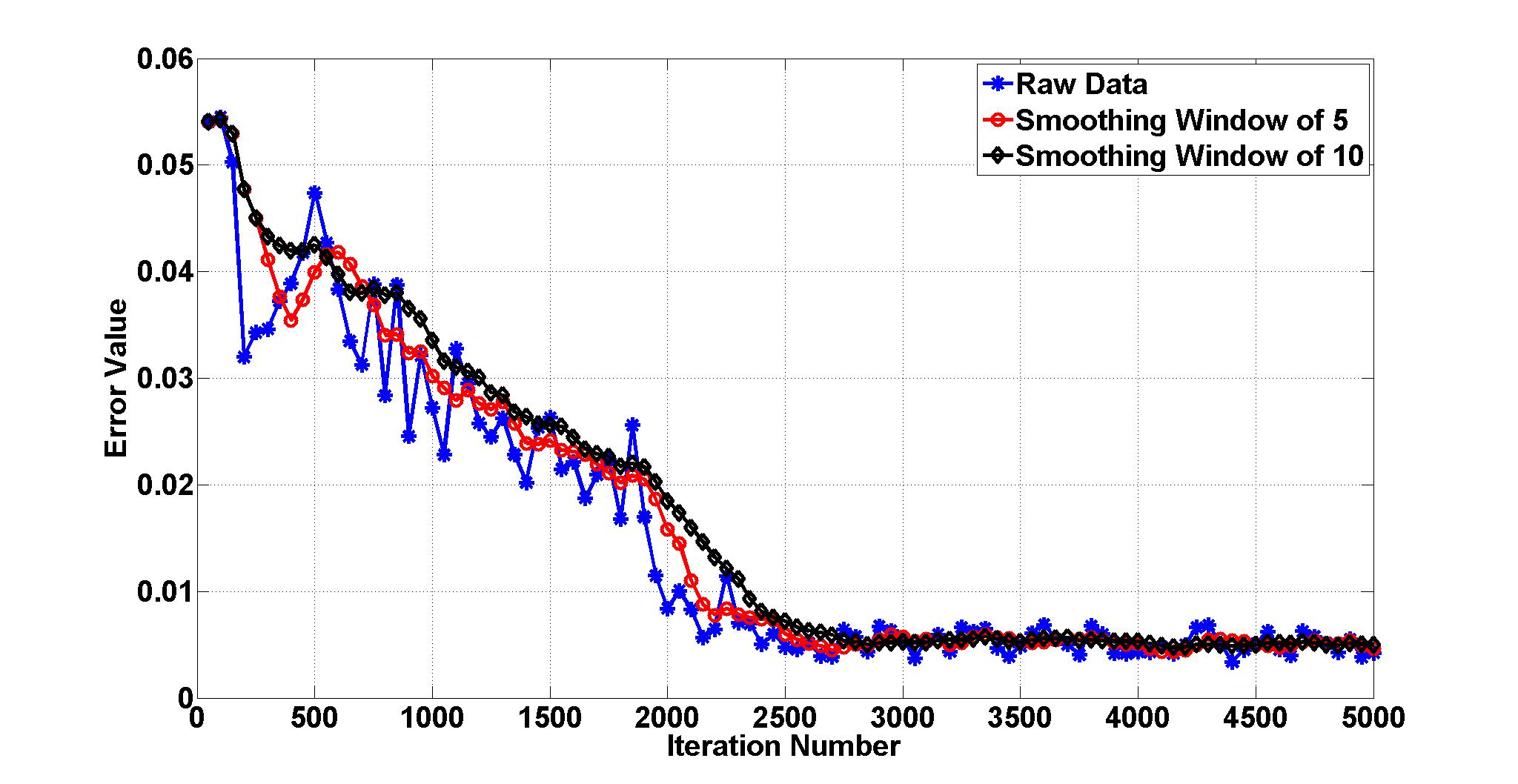}  
 \end{subfigure}
 \caption{The MSE error graph of the validation portfolio as a function of iteration number and smoothing window value.}
 \label{fig:error-smoothing-window}
\end{figure}

We use data smoothing and polynomial fitting to extract the major u-shape trend in the MSE graph and hence find stopping events. In order to 
find a good choice for the smoothing window, we start with a small value of the smoothing window and calculate the smoothed curve. 
If the small peaks and valleys of the orignial curve are supressed and big peaks and big valleys of the original curve are significantly damped, 
then we choose that value of the smoothing window as our final choice for the smoothing window. For example, in Figure \ref{fig:error-smoothing-window}, 
the smoothed curve with a smoothing window of $5$ still has a big valley around iteration number of 400. However the valley is dampened in the smoothed 
graph resulting from smoothing window of $10$.    

The primary goal of the polynomial fitting is to find the u-shaped trend in the graph so that we can detect the stopping event. 
The u-shaped trend therefore suggests that the polynomial should go to infinity as its argument goes to either plus or minus infinity. 
Therefore, the degree of the polynomial should be even. Since we are only interested in detecting a u-shaped trend, it is sufficient 
to use polynomials of low degree ($\leq 10$). High degree polynomials overfit the data and they can't detect a slowly increasing trend such as 
the one in Figure \ref{fig:error-smoothing-window} after iteration 2500. On the other hand, a simple polynomial of degree 2 does not always work well. 
A quadratic polynomial on a MSE graph similar to Figure \ref{fig:error-smoothing-window} falsely detects a u-shape trend in the big valley between 
iteration numbers 0 and 500. However a polynomial of degree 4 or higher will not make such a detection. Because we smooth the data before we fit any 
polynomials and we choose our learning parameter such that we expect an initial decreasing trend, we suggest polynomials of degree $4, 6$ or $8$ 
to be used to fit the data to find u-shaped trends. 

Finally for the value of window length to detect that we have reached the minimum, we choose a value of $W$ such that the number of iterations in the window ($W \times I$) 
is big enough (around a hundred iterations) that we can confidently say the graph of the MSE error has reached a minimum value and started to increase in value (an 
increasing trend). Notice that the window length should not be too big so that we can start the search in the local neighborhood and minimize the training time.